\documentclass[iop]{emulateapj}
\defcitealias{kempner2004}{KD04}
\defcitealias{kempner2003}{KSM03}
\defcitealias{owers2011}{O11}

\usepackage{epsfig}
\usepackage{subfigure}
\usepackage{graphics}

\newcommand{\chan}{{\it Chandra}}

\newcommand{\kms}{$\,\rm{km\,s^{-1}}$}
\newcommand{\msolar}{$\rm{M}_{\odot}$}
\newcommand{\neout}{n_{\rm {e,\,out}}}
\newcommand{\nein}{n_{\rm {e,\,in}}}
\newcommand{\ktout}{kT_{\rm{out}}}
\newcommand{\ktin}{kT_{\rm{in}}}
\shorttitle{}
\shortauthors{Owers et al.}
\begin{document}
\title{A Merger Shock in Abell~2034.}
\author{Matt S. Owers\altaffilmark{1,5}, Paul E.J. Nulsen\altaffilmark{2}, Warrick J. Couch\altaffilmark{1,3}, Cheng-Jiun Ma\altaffilmark{2,4}, Laurence P. David\altaffilmark{2}, William R. Forman\altaffilmark{2}, Andrew M. Hopkins\altaffilmark{1}, Christine Jones\altaffilmark{2}, Reinout J. van Weeren\altaffilmark{2}}

\altaffiltext{1}{Australian Astronomical Observatory, PO Box 915, North Ryde, NSW 1670, Australia}
\altaffiltext{2}{Harvard Smithsonian Center for Astrophysics, 60 Garden Street, Cambridge, MA 02138, USA}
\altaffiltext{3}{Center for Astrophysics and Supercomputing, Swinburne University of Technology, Hawthorn, VIC 3122, Australia}
\altaffiltext{4}{Department of Physics \& Astronomy, University of Waterloo, 200 University Ave. W., Waterloo, Ontario, N2L 3G1, Canada}
\altaffiltext{5}{Australian Research Council Super Science Fellow; Email: matthew.owers@aao.gov.au}

\begin{abstract}
We present a 250\,ks \chan\ observation of the cluster merger A2034 with the aim of understanding the nature of a sharp edge previously characterized as a cold front. The new data reveal that the edge is coherent over a larger opening angle and is significantly more bow-shock-shaped than previously thought. Within $\sim 27\degr$ about the axis of symmetry of the edge the density, temperature and pressure drop abruptly by factors of $1.83^{+0.09}_{-0.08}$, $1.85^{+0.41}_{-0.41}$ and $3.4^{+0.8}_{-0.7}$, respectively. This is inconsistent with the pressure equilibrium expected of a cold front and we conclude that the edge is a shock front. We measure a Mach number $M = 1.59^{+0.06}_{-0.07}$ and corresponding shock velocity $v_{\rm shock}\simeq 2057$\kms. Using spectra collected at the MMT with the Hectospec multi-object spectrograph we identify 328 spectroscopically confirmed cluster members. Significantly, we find a local peak in the projected galaxy density associated with a bright cluster galaxy which is located just ahead of the nose of the shock. The data are consistent with a merger viewed within $\sim 23\degr$ of the plane of the sky. The merging subclusters are now moving apart along a north-south axis approximately $0.3\,$Gyr after a small impact parameter core passage. The gas core of the secondary subcluster, which was driving the shock, appears to have been disrupted by the merger. Without a driving ``piston'' we speculate that the shock is dying. Finally, we propose that the diffuse radio emission near the shock is due to the revival of pre-existing radio plasma which has been overrun by the shock.
\end{abstract}


\section{Introduction}

Cluster mergers are natural outcomes of the hierarchical nature of large scale structure formation. Major mergers involving clusters of a similar mass are rare, although the impact on the cluster constituents can be significant given the extreme kinetic energies of up to $10^{64}\,$erg \citep{markevitch1999}. Approximately $10\%$ of this energy is dissipated in the collisional intracluster medium (ICM) through shocks, turbulence and compression which produce spectacular observable effects at X-ray wavelengths \citep{markevitch2007,sarazin2008}. Observations of cluster mergers have proven valuable in constraining the nature of dark matter \citep[e.g., in the Bullet cluster;][]{clowe2006,randall2008b}, the cosmic ray content of clusters \citep{feretti2012}, the intracluster magnetic field strength \citep{vikhlinin2001a}, and for understanding the physics of the ICM \citep[][]{ettori2000,vikhlinin2001b, russell2012,roediger2013}. These studies show that deep multiwavelength observations of cluster mergers are critical if we are to understand the impact of large scale structure formation on the cluster constituents.

Early X-ray observations of cluster mergers using the {\it ROSAT} and ASCA satellites revealed a hot, apparently shock-heated ICM in regions affected by merger activity \citep{henriksen1996,donelly1998,markevitch1999}. However, the spatial resolution was not sufficient to see the surface brightness edge expected due to the compression of gas at the shock front. The launch of the \chan\ X-ray observatory with its sub-arcsecond resolution and excellent sensitivity was expected to reveal many shock edges in merging clusters. In fact, while \chan\ has uncovered a number of merger-related edges, the vast majority are cold fronts \citep[][]{markevitch2000,vikhlinin2002,owers2009c} and only a handful of shock fronts have been directly observed as sharp edges in \chan\ images \citep[in the Bullet, A520, A2146, A2744 and A754;][]{markevitch2002, markevitch2005, russell2010,russell2012, owers2011a, marcario2011}. The rarity of shock fronts is due to a number of factors which conspire against them being easily observable as edges in X-ray images. First, the edge created by the compression of gas at the shock front is best observed when the shock motion is very close to the plane of the sky. Second, shocks are most prominent in the central regions of clusters where the X-ray surface brightness is high. This constrains observability to the relatively short core passage phase of a merger. Third, merger velocities are of the order of a few thousand km/s so the shocks have Mach numbers $M \lesssim 3$. This means that the temperature contrast across the edge can be low and, since temperature measurements are limited by the available counts, high quality observations are required to accurately measure the temperature jumps required to characterize edges as shocks.

Here we present new and significantly deeper \chan\ observations of the merging cluster Abell~2034 (hereafter A2034). The previous $54\,$ks \chan\ observation presented by \citeauthor{kempner2003}~(2003; hereafter \citetalias{kempner2003}) revealed an edge to the north of the cluster which was interpreted as a cold front, although it was noted that the temperature contrast across the edge was not as strong as in other cold fronts. Cold fronts are contact discontinuities which occur at the interface of low-entropy gas moving through a higher-entropy medium. By definition, the pressure across a contact discontinuity is continuous. \citet{owers2009c} subsequently excluded A2034 from their sample of cold front clusters since the pressure across the edge did not appear to be continuous. The properties of the edge indicated that it may be a shock, although the data at hand were not of sufficient depth to conclusively show this. Interestingly, radio observations of A2034 show evidence for diffuse emission near the northern edge which was tentatively classified as a radio relic \citep[][]{kempner2001,vanweeren2011} although \citet{rudnick2009} classify the diffuse radio emission in A2034 as a halo. The emission from radio halos and relics is a result of the interaction of a population of relativistic electrons with the cluster magnetic field which results in synchrotron radiation at radio wavelengths. While the favored model for the production of relativistic electrons producing the large scale ($\gtrsim 1\,$Mpc) radio halos is reacceleration of mildly relativistic electrons due to merger driven turbulence \citep{brunetti2008}, it is likely that shocks provide the mechanism by which radio relics are formed \citep{feretti2012}. Models for shock-induced radio relics include the direct acceleration of thermal electrons \citep{ensslin1998}, the reacceleration of preexisting mildly relativistic electrons \citep{markevitch2005} and the revival of fossil radio plasma by adiabatic compression from the shock \citep{ensslin2001}. The existence of a radio relic near the northern edge may therefore be taken as indirect evidence that the edge is a shock front.

The available $250\,$ks of \chan\ data allow us to show that the edge in A2034 is a shock front. The characterization of this edge as a shock, combined with the information provided from our optical spectroscopy taken with the MMT-Hectospec instrument, allow us to make a more precise determination of the merger history in A2034.  The paper is structured as follows: In Section~\ref{data} we describe the \chan\ and MMT data to be used in our analysis. In Section~\ref{analysis} we present images and temperature maps of the ICM, surface brightness and temperature profiles across the edge, define spectroscopically confirmed cluster membership and determine the spatial distribution of cluster members. In Sections~\ref{discussion} and \ref{conclusion} we discuss our results and present our conclusions. Throughout the paper, we assume a standard $\Lambda$CDM cosmology with $\Omega_{\rm M}=0.3$, $\Omega_{\Lambda}=0.7$ and $H_0=70$\kms\,${\rm {Mpc}^{-1}}$. The cluster redshift determined from our sample of confirmed cluster members is $z=0.1132$ (Section~\ref{membership}) which, given the assumed cosmology, means $1\arcsec = 2.06\,$kpc.

\section{Observations and data processing}\label{data}

\subsection{Chandra Data}\label{chandra_data}
Details of the \chan\ observations used in this paper are summarized in Table~\ref{xray_obs}. All \chan\ observations were taken on the ACIS-I chip array. The 54\,ks \chan\ observation used in \citetalias{kempner2003} (\chan\ ObsID 2204) was taken on 2001 May. The newer 200\,ks exposure was split into four separate observations (ObsIDs 12885, 12886, 13192 and 13193) taken during 2010 November. The aimpoint of the new observations was carefully chosen to ensure that none of the chip gaps obscure critical regions near the edge discussed in \citetalias{kempner2003}. The \chan\ data were reprocessed using the CHANDRA\_REPRO script within the CIAO software package \citep[version 4.4][]{fruscione2006}. The script applies the latest calibrations to the data (CALDB 4.5.6), creates an observation-specific bad pixel file by identifying hot pixels and events associated with cosmic rays (utilizing VFAINT observation mode) and filters the event list to include only events with {\it ASCA} grades 0, 2, 3, 4, and 6. The DEFLARE script is then used to identify and filter periods contaminated by background flares. Table~\ref{xray_obs} lists the total and flare-filtered exposure times and it shows that our observations were not affected by significant background flares. For the imaging analyses, exposure maps which account for the effects of vignetting, quantum efficiency (QE), QE non-uniformity, bad pixels, dithering, and effective area were produced using standard CIAO procedures\footnote{cxc.harvard.edu/ciao/threads/expmap\_acis\_multi/}. The energy dependence of the effective area is accounted for by computing a weighted instrument map with the SHERPA {\sf make\_instmap\_weights} script using an absorbed MEKAL spectral model with $N_{\rm H}=1.58\times 10^{20} {\rm cm}^{-2}$ \citep{dickey1990}, the average cluster values of $kT = 7.9$\,keV and  abundance 0.29 times solar \citepalias{kempner2003} and $z=0.113$.

\begin{deluxetable}{ccccc}
  \tablecaption{Summary of the five \chan\ X-ray pointings. \label{xray_obs}} 
  \tablecolumns{5}
\tablehead {\colhead{ObsIDs}   & \colhead{R.A.} & \colhead{decl.} & \colhead{$T_{\rm exp}$} &  \colhead{Cleaned $T_{\rm exp}$}\\
& & &  \colhead{(ks)} & \colhead{(ks)}}\\
\startdata
\dataset[ADS/Sa.CXO#obs/02204]{2204} &  15:10:11.71  &   +33:29:11.79  & 53.95 & 53.15 \\
\dataset[ADS/Sa.CXO#obs/12885]{12885} &  15:10:13.40 & +33:30:43.00	 & 81.2 & 80.69 \\
\dataset[ADS/Sa.CXO#obs/12886]{12886} &  15:10:13.40 & +33:30:43.00 & 91.3 & 91.3 \\
\dataset[ADS/Sa.CXO#obs/13192]{13192} &  15:10:13.40 & +33:30:43.00 & 16.83 & 16.33 \\
\dataset[ADS/Sa.CXO#obs/13193]{13193} &   15:10:13.40 & +33:30:43.00 & 7.67 & 7.67 
\enddata
\end{deluxetable}

Background subtraction for both imaging and spectroscopic analyses was performed using the period D and F blank sky backgrounds\footnote{cxc.harvard.edu/contrib/maxim/acisbg/}$^,$\footnote{cxc.harvard.edu/caldb/downloads/Release\_notes/supporting/\\README\_ACIS\_BKGRND\_GROUPF.txt} which were processed in the same manner as the observations. The blank sky backgrounds were reprojected to match the tangent point of the observations, and were normalized to match the $9-12\,$keV counts in the observations using the source-free I0 and I2 chips. For the imaging analysis, we also made use of Maxim Markevitch's {\sf make\_readout\_bg}\footnote{cxc.cfa.harvard.edu/contrib/maxim/make\_readout\_bg} script to produce a readout background to be subtracted as described in \citet{markevitch2000}. This accounts for the readout streak produced by a bright point source at R.A.=15:10:41.2, decl.=33:35:05.1 which is associated with a cluster member galaxy. These readout streaks were masked during spectroscopic analyses.

\subsection{MMT Hectospec data}\label{hecto_data}

We observed four fiber configurations over the period 2011 February to April centered on A2034 using the 300-fiber Hectospec multi-object spectrograph on the 6.5m MMT \citep[see ][for instrument details]{fabricant2005}. Targets for the observations were selected using catalogs downloaded from the Sloan Digital Sky Survey DR7 \citep{abazajian2009}. We only included those objects which were within a radius of $30\arcmin = 3.7\,$Mpc of the central brightest cluster galaxy (BCG; R.A.=15:10:11.7, decl. = 33:29:11.5) and which were classified by the SDSS pipeline as galaxies. Galaxies which have colors that are redder than the cluster red-sequence are generally not cluster members. Therefore, we only included galaxies which have $B-R$ colours placing them on, or blueward of, the cluster red-sequence which was defined using existing spectroscopically confirmed cluster member galaxies from SDSS spectroscopy. We excluded objects with existing redshifts which place them well outside the redshift range of potential cluster members. 

The four configurations were split into two bright and two faint configurations where the magnitude limits were $R < 19.5$ and $r < 20.3$, respectively. The bright configurations were observed for $3\times 20$\,minute exposures in generally good conditions with seeing FWHM=$0.73\arcsec - 0.94\arcsec$. The faint configurations had one set of $6\times 20$\,minute exposures in relatively poor seeing (FWHM=$1.78\arcsec$) and another with $(5\times 20 + 1\times 480)$\,minute exposures in $0.81\arcsec$ seeing. The observations were performed using the 270 groove mm$^{-1}$ grating which provided coverage of the wavelength interval $3500-10000\,$\AA\, at $\sim 6\,$\AA\, resolution. Around 30 fibers per configuration were allocated to blank sky regions for the purpose of sky subtraction. The spectra were reduced at the Smithsonian Astrophysical Observatory Telescope Data Center\footnote{tdc-www.harvard.edu} (TDC) using the SPECROAD\footnote{tdc-www.harvard.edu/instruments/hectospec/specroad.html} pipeline \citep{mink2007}. Redshifts were also determined at the TDC using the IRAF cross-correlation XCSAO software \citep{Kurtz1992} and each spectrum was assigned a redshift quality of ``Q'' for a reliable redshift, ``?'' for questionable, and ``X'' for a bad redshift measurement. These observations provided 736 quality ``Q'' redshift measurements for extragalactic objects. 

\section{Results and Analysis}\label{analysis}

\subsection{X-ray image}\label{chan_image}

\begin{figure}
\includegraphics[width=0.45\textwidth]{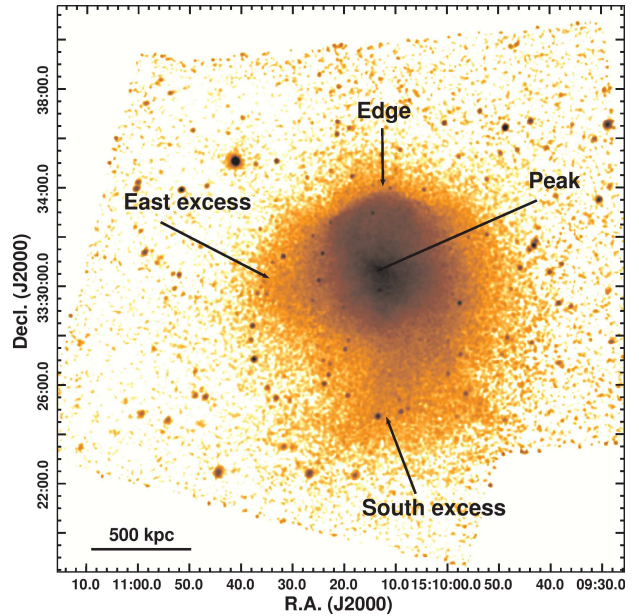}
\caption{Combined, exposure-corrected and background-subtracted \chan\, image in the $0.5-7.0\,$keV energy band. The pixels are $1.968\arcsec$ on a side and the image has been smoothed with a Gaussian kernel with FWHM=$6\arcsec$. Note the sharp edge to the north which we argue is a Mach $1.59$ shock. \label{chan_img}}
\end{figure}

In Figure~\ref{chan_img}, we show a background-subtracted, exposure-corrected mosaic of the \chan\ pointings. The most prominent feature is the sharp edge $\sim 190\arcsec \,(390\,$kpc) to the north of the peak in the X-ray surface brightness. This is the edge that was interpreted as being due to a cold front by \citetalias{kempner2003}, but called into question as such by \citet{owers2009c}. The deeper data reveal that the edge is well defined over a larger opening angle than that seen in \citetalias{kempner2003} and has a morphology resembling a bow-shock-cone seen in projection. The axis of symmetry (hereafter referred to as the ``nose'') is approximately due north of the X-ray brightness peak and corresponds to the central sector in Figure~9a of \citetalias{kempner2003}. Under their assumption of a circular front, the nose of the shock cone was interpreted as an excess of emission located ahead of the edge by \citetalias{kempner2003}. It is possible that their analysis was hindered by the position of a chip gap in ObsID 2204 over the western portion of the edge which significantly reduced the exposure in that region. Addressing the physical nature of the edge is the focus of this paper, and we will revisit the thermodynamic properties of the edge in more detail in Section~\ref{profiles}. 

\begin{figure}
\includegraphics[width=0.45\textwidth]{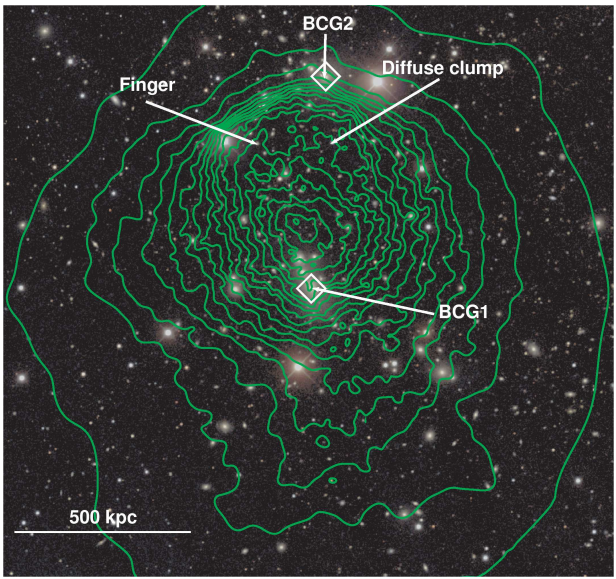}
\caption{SDSS RGB image using the g-, r- and i-bands. Contours from a background-subtracted, exposure-corrected and adaptively smoothed \chan\ image are overlaid. The smoothing kernel size is set such that the S/N per pixel is $\sim 15$. There are 20 contours starting at $5.9 \times 10^{-9}$ and have ASINH spacing to the maximum value of $6.2 \times 10^{-7}$. \label{opt_chan}}
\end{figure}

As noted by \citetalias{kempner2003}, to the south of the X-ray peak there is an extended region of low surface brightness emission which is $\sim 4$\,\arcmin\ long. Our deeper data show that this low surface brightness emission is elongated with the axis of elongation pointing just east of south. Based on the fairly regular, diffuse morphology revealed by their shallower data, \citetalias{kempner2003} offered three explanations for this excess. Two of these explanations centred around the southern excess being associated with a smaller subcluster which is merging with A2034. The third and most favored explanation was that the south excess was due to the emission associated with a background cluster seen in projection through A2034. The highly elongated morphology, along with the lack of any clear galaxy association in the deeper SDSS images (Figure~\ref{opt_chan}) indicate that the background cluster interpretation may be incorrect.

In Figure~\ref{opt_chan} we overlay contours from an adaptively smoothed \chan\ image onto an SDSS RGB image. The peak in the X-ray emission is 172\,kpc to the north of the BCG (hereafter, BCG1), while there does appear to be a less-prominent peak in the X-ray emission coincident with BCG1. Just north of the nose of the edge is another bright cluster member which we have labeled BCG2. Some $\sim 200$\,kpc south of the nose is a diffuse clump of emission, while a finger of emission extends from the eastern portion of the edge toward the south. The deeper data also reveal a low surface brightness asymmetry $\sim 400\,$kpc east of the peak in the X-ray brightness.

\subsection{Temperature and Hardness ratio Maps}\label{tempmaps}

For the purpose of understanding the spectral properties of the ICM in A2034, we construct temperature and hardness ratio maps. The hardness ratio, defined here as the ratio of the flux in the $2-7$\,keV and $0.5-7$\,keV bands, does not produce a direct measure of the gas temperature. However, for a given number of source counts a hardness ratio map provides a simpler, higher signal-to-noise, model-independent diagnostic of the spectral properties in a region when compared with a single thermal component temperature map. Moreover, a hardness ratio map may reveal regions where a single component thermal model does not adequately describe the spectral properties of the ICM. Therefore, the hardness ratio and temperature maps can provide a complementary, quasi-independent, spatially resolved probe of the ICM spectral properties.

\begin{figure*}
\includegraphics[width=0.49\textwidth]{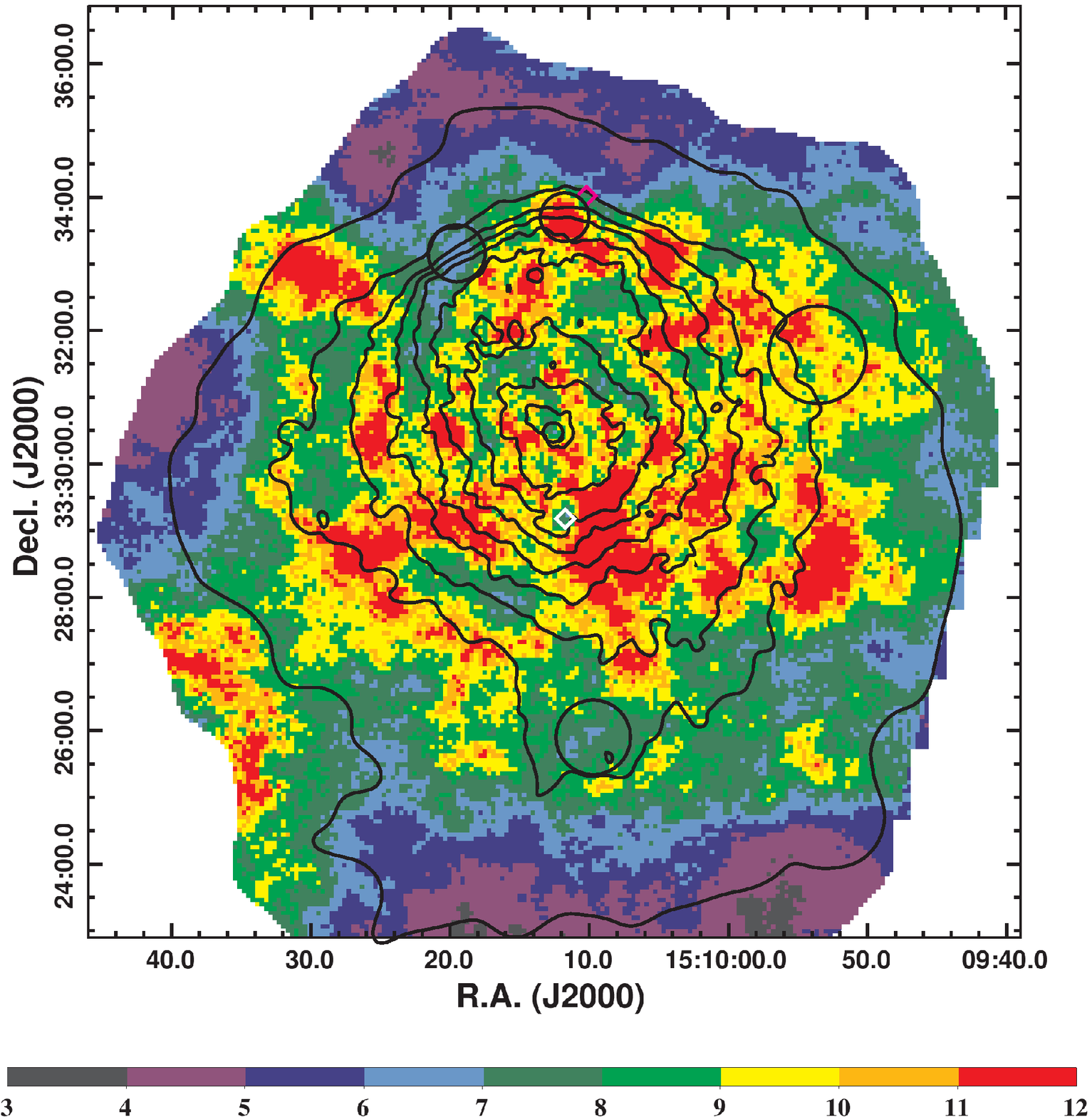}
\includegraphics[width=0.50\textwidth]{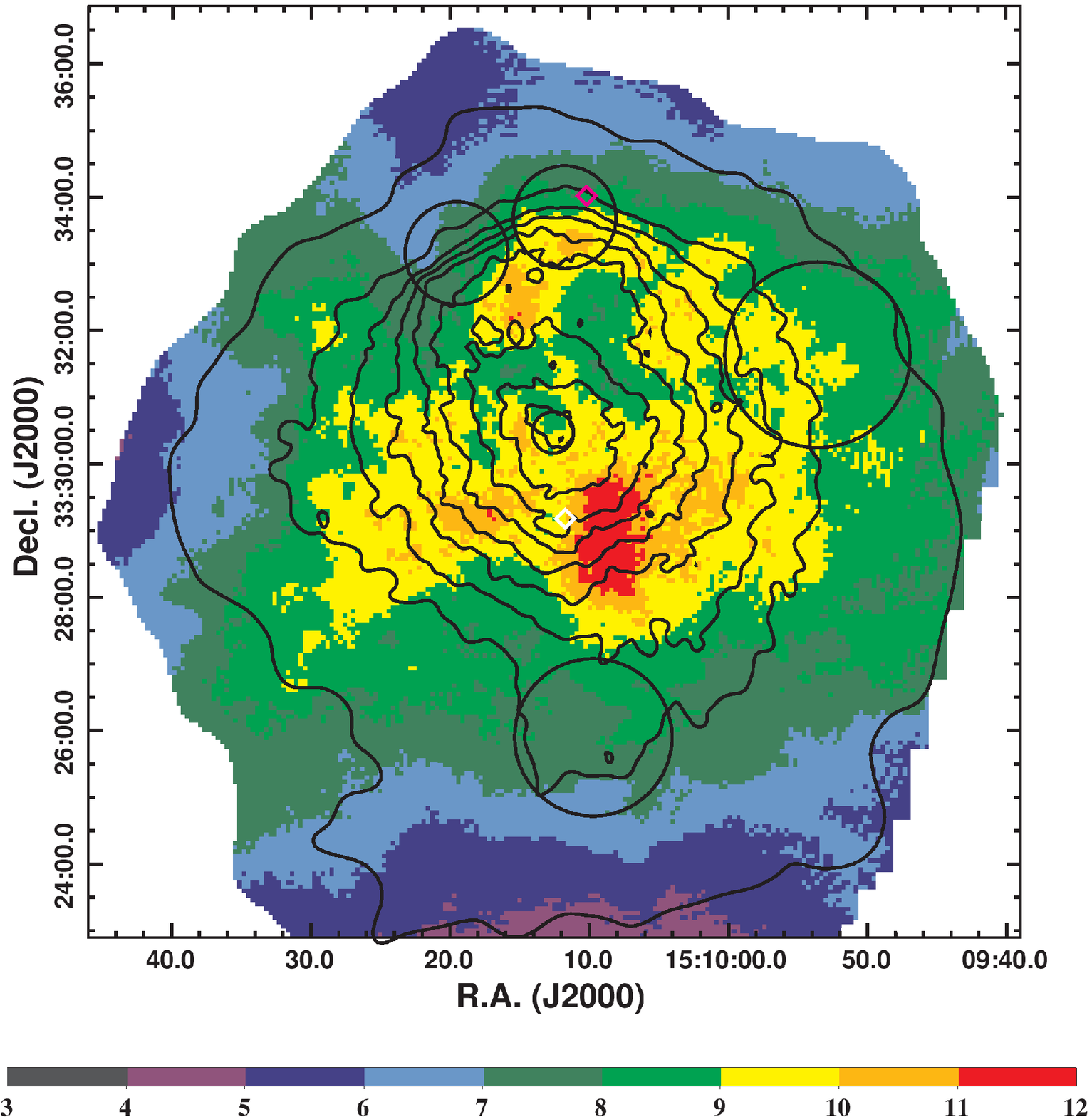}\\
\includegraphics[width=0.49\textwidth]{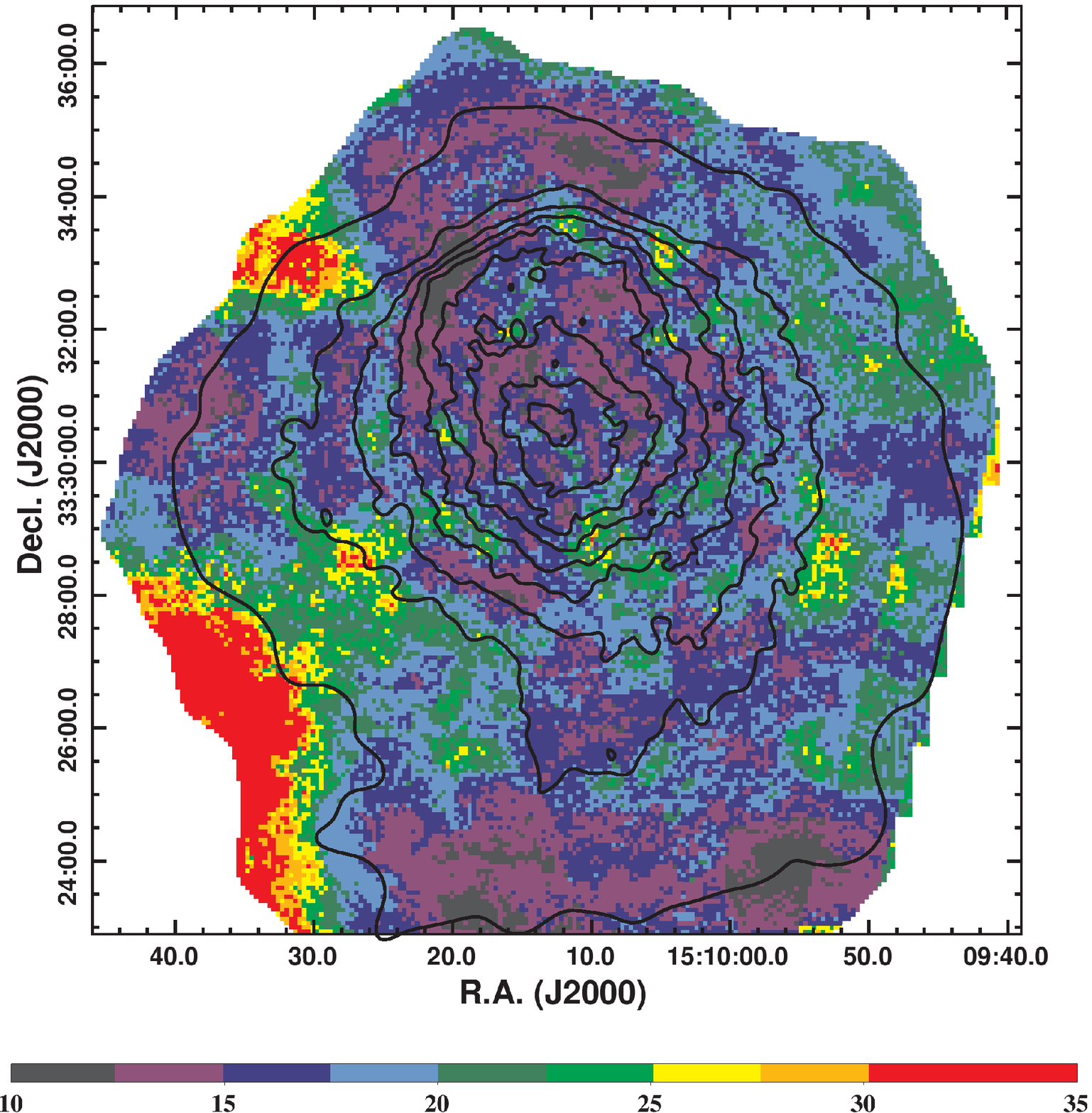}
\includegraphics[width=0.50\textwidth]{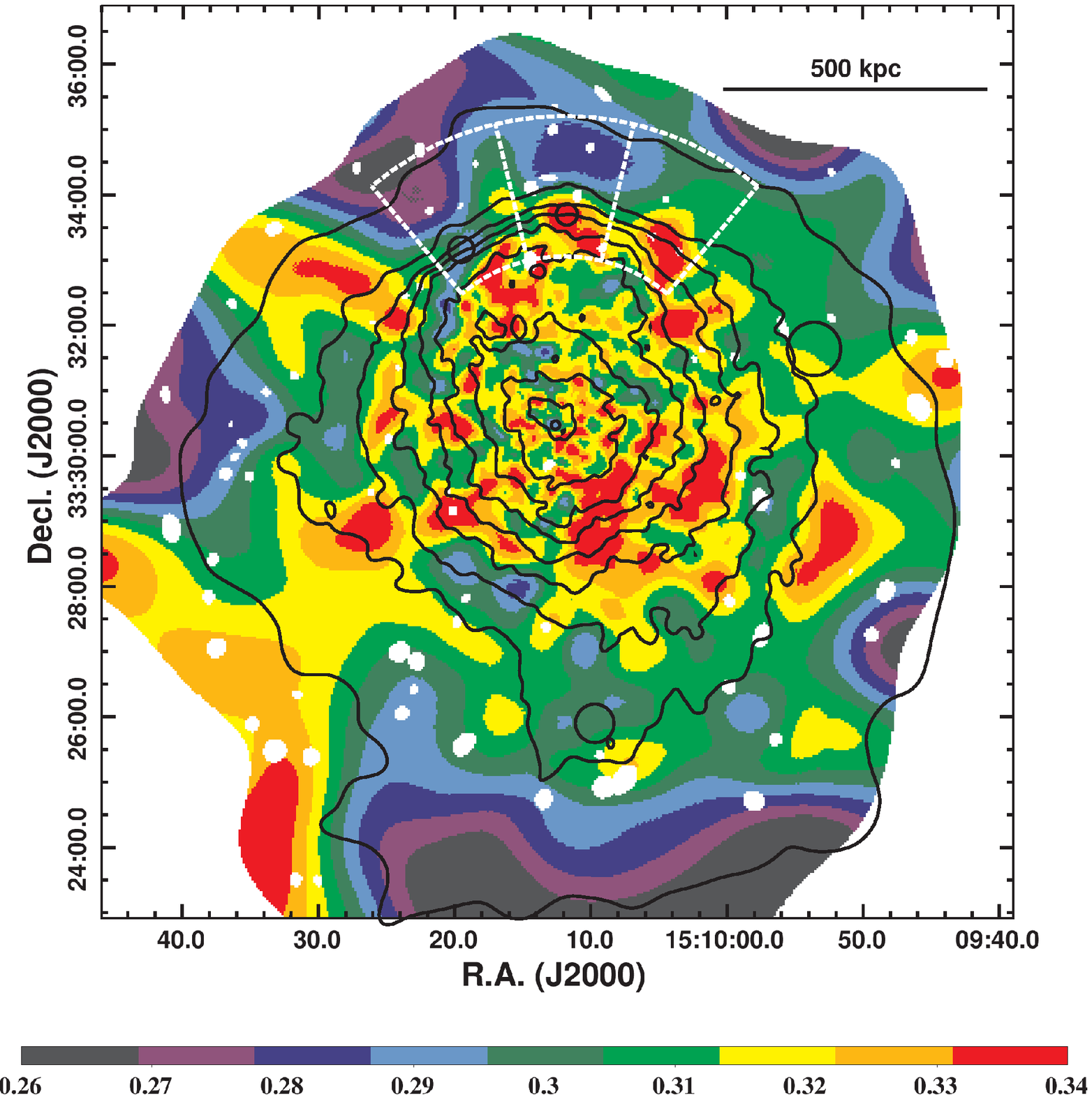}\\

\caption{Upper Left panel: High resolution temperature map with surface brightness contours overlaid in black. The temperature at each pixel is obtained by fitting an absorbed MEKAL model to a spectrum which is extracted from an adaptively sized circular region with radius set such that the spectrum has 2000 background-subtracted counts. The black circles show the size of the extraction regions at their position. The diamond points show the location of the BCG1 (white diamond point) and BCG2 located ahead of the nose of the edge (magenta diamond point). Note the high temperature at the nose, and the rapid decline in temperature moving to the fainter side of the edge. This indicates that the edge is a shock. The units of the colour bar are keV. 
Upper Right panel: A lower resolution, 8000 count, temperature map. Note the qualitative agreement with the higher resolution 2000 count map on large scales, although many smaller scale features are smoothed out. 
Lower left panel: Upper confidence range for the 2000 count temperature map expressed as a fraction of the best fit temperature. The confidence intervals are skewed about the best fit, so that the lower confidence range is typically a factor of $\sim 1.3$ smaller than the upper range.  Values on the color bar are expressed as percentages.
Lower right panel: Hardness ratio map derived by taking the ratio of the adaptively smoothed 2-7\,keV and 0.5-7\,keV band images. The smoothing length is set such that the relative uncertainty on the hardness ratio is $\sim 5\%$ and the black circles show the $1\sigma$ Gaussian smoothing length at that position. The dashed white annular sectors show the regions from which the surface brightness and temperature profiles were extracted in Section~\ref{profiles}. Note the finger of soft emission running from the eastern portion of the edge to the south. This soft emission does not follow the surface brightness contours which define the edge. \label{tmap}}
\end{figure*}

The temperature maps are generated in a similar manner to that described in \citet{randall2008} and are shown in the top left and top right panels of Figure~\ref{tmap}. Briefly, each temperature map has pixels with a constant size ($3.94\arcsec \times 3.94\arcsec$) and spectra for each pixel are extracted from a circular region with radius set such that the region contains a constant number of background subtracted 0.5-7\,keV source counts. We produce a high spatial resolution map with 2000 source counts per extraction region, and a lower spatial resolution map with 8000 source counts per extraction region. The higher resolution temperature map provides a more accurate picture where there are abrupt changes in temperature, such as near shock and cold fronts. The lower resolution temperature map provides more accurate temperatures, particularly for the higher temperature regions, although it significantly smears out sharp gradients in temperature. For both maps, but more so the 8000 count map, in the low surface brightness regions the pixel values are correlated over large distances. To demonstrate this effect, we show extraction region sizes as black circles at five locations in the top left and top right panels of Figure~\ref{tmap}.

Redistribution Matrix Files (RMFs) and Ancilliary Response Files (ARFs) are generated with the CIAO {\sf mkacisrmf} and {\sf mkwarf} tools, respectively. Because the RMFs can take several minutes to process, and because there are $\sim 200,000$ spectra to process, we extract the RMFs and ARFs on a more coarsely binned grid with $15.74\arcsec \times 15.74 \arcsec$ pixels. The size of the extraction regions are generally much larger than the pixel size of this coarser grid, so the impact on the temperature measurements is small and well within the measurement uncertainties. The five spectra per pixel are fitted simultaneously in XSPEC \citep[Version 12.7.1][]{arnaud1996} with an absorbed MEKAL model \citep{mewe1985, mewe1986, Kaastra1992,liedahl1995}. During the fitting the temperature is left as a free parameter which is tied across the spectra and the abundance is fixed at the global value $Z=0.33$ which is measured with respect to solar values \citep[using ratios from][]{anders1989}. The normalizations are also left as free parameters, however for observations with similar pointings the values are tied together. Spectra are constrained to the energy range 0.7-7\,keV, binned to contain at least one count per energy bin, and the modified Cash statistic, WSTAT, is minimized\footnote{heasarc.gsfc.nasa.gov/docs/xanadu/xspec/manual/\\XSappendixStatistics.html}.

The lower left panel of Figure~\ref{tmap} shows the upper confidence interval for the higher resolution, 2000 count, temperature map expressed as a fraction of the best fit temperature. The confidence intervals are measured at the $68\%$ level (corresponding to $\Delta (WSTAT) =1$) and are determined for a single parameter of interest. The confidence intervals are skewed about the best fit so that the upper confidence interval is generally a factor of $\sim 1.3$ higher than the lower confidence interval. Where the X-ray surface brightness is high, the upper confidence intervals are $10-15\%$ and $20-25\%$ of the best fit temperature for regions which have $kT < 10$\,keV and $kT > 11$\,keV, respectively. In the lower surface brightness regions, where the background flux begins to dominate, the relative uncertainties become much larger ($>30\%$) where $kT > 11$\,keV. Given these uncertainties, regions with $kT < 8\,$keV are cooler than regions with $kT > 10\,$keV with $\gtrsim 90\%$ confidence.

Comparing the higher and lower resolution temperature maps, there is qualitative agreement on large scales. Clearly the distribution of temperatures is less patchy for the lower resolution, 8000 count, map and this is due to two main reasons. First, hot regions where $kT \gtrsim 11\,$keV are less well constrained in the higher resolution, 2000 count, map and excursions to much higher temperatures on smaller scales may simply be due to statistical fluctuations which are not significant. The relative uncertainties on the lower resolution map are generally $<10\%$ and so the temperature map is considerably less noisy. Second, the large aperture required to collect 8000 counts smooths out real fluctuations in temperature which occur on scales smaller than the aperture. For example, the extraction region near the sharp edge to the north is roughly twice the diameter in the lower resolution map when compared with the higher resolution map (see the black circles in the top panels of Figure~\ref{tmap}). Therefore, the temperature measurement on the brighter side of the edge is significantly contaminated by flux from the fainter side of the edge, washing out the sharp temperature gradient which is significant in the higher resolution map. For the remainder of this section, we focus our discussion around significant features seen in the higher resolution, 2000 count, map.

The lower right panel of Figure~\ref{tmap} shows the hardness ratio map. The map is produced by smoothing the background-subtracted and exposure-corrected 2-7\,keV and 0.5-7\,keV band images with a Gaussian kernel with an adaptive smoothing length. Rather than choosing an adaptive smoothing width based on one of the images \citep[e.g., as in ][]{henning2009}, we follow a similar method to that used in \citet{sanders2001}. That is, we set the smoothing length such that the hardness ratio at each pixel has a relative uncertainty of $\sim 5\%$. Comparison of the hardness ratio and temperature maps show that high temperature regions in the high resolution temperature map clearly map regions which are significantly harder in the hardness ratio map. This provides confidence that the regions with high temperature values in the 2000 count temperature map are regions with significantly harder, and likely hotter, emission. However, near the eastern portion of the edge there some differences between the high resolution temperature map and the hardness ratio map which we discuss below.


The distribution of temperatures is patchy and complicated and is broadly consistent with the map presented in \citetalias{kempner2003}, although the deeper observations reveal several features associated with the northern edge which were not previously identified. The region corresponding to the nose of the northern edge has $kT\gtrsim 11$\,keV and is significantly hotter than the ICM lying further north of the nose, which has $kT\simeq 6$\,keV. This hotspot is highlighted by a black circle the radius of which shows the size of the extraction region there. While \citetalias{kempner2003} discuss a hot spot in the vicinity of the nose, it appears that the peak temperature in the hotspot seen in their Figure~7 lies somewhat further south and away from the nose. We confirm that the region discussed in \citetalias{kempner2003} is in fact hotter than its surrounds, but stress that it is separate from the hotspot detected at the nose. The sector of the edge lying to the west of the nose also shows a strong gradient in temperature in the sense that the temperature drops sharply going from the brighter side of the edge ($\sim 9$\,keV) to the fainter side ($\sim 5-6$\,keV). We note a second hot spot with $kT\gtrsim 12$\,keV at the westernmost point of the edge. Thus, the nose and western sectors of the edge appear to harbour thermal properties which are more consistent with the edge being a shock front rather than a cold front.

In defining the position of the edge, \citetalias{kempner2003} appear to have focused on the sector just east of the nose. Figure~\ref{tmap} shows that the thermal properties of this eastern sector are somewhat abnormal. Near to the nose the temperature distribution is shock-like; inside the edge the temperature is $\simeq 9$\,keV and drops abruptly to the north to $\simeq 7$\,keV although there are several patches of hotter $\sim 8$\,keV gas further ahead of the edge. These hotter $\simeq 8$\,keV patches are also seen as a region of hard emission in the hardness ratio map. Following the edge further southeast, the temperature on the bright side of the edge drops abruptly to $6.5-7.5$\,keV, similar to the temperature just outside the edge there. The constant temperature going from the inside to the outside of the edge does not appear to be consistent with the behavior expected of a shock or cold front. The hardness ratio map shows a finger of emission which starts at the edge and extends to the south. This finger is significantly softer than the surrounding ICM. The hardness ratio map has better spatial resolution than the temperature map, and it shows that this soft finger does not follow the surface brightness contours which trace the edge. The nature of this soft finger of emission is unclear, but it is likely responsible for the odd behaviour observed in the temperature map near the eastern portion of the edge. Since projection effects and cross contamination of emission from either side of the edge may wash out temperature gradients at the edge, we will investigate the profile across the edge further in Section~\ref{profiles} using well-defined regions inside and outside the edge.

Aside from the edge, there are several other features of note in the temperature map. As noted in \citetalias{kempner2003}, the temperature at the peak in the X-ray surface brightness is cooler than its surrounds. To the south of the peak in the X-ray surface brightness is a band of hot $\gtrsim 11$\,keV gas that extends $\sim 750$\,kpc along an east-west axis \citep[similar to the hot region seen just west of the main cluster in the Bullet;][]{owers2009c}. Embedded within this hot band is a region of enhanced surface brightness and cooler $kT\simeq 7.5$\,keV gas which is spatially coincident with BCG1 (open white diamond symbol in the left panel of Figure~\ref{tmap}). Further south, the temperature maps show that the extended region of low surface brightness emission discussed in Section~\ref{chan_image} harbors cooler $\sim 6.5-7.5$\,keV gas. At large radii, the temperature of the gas is generally much cooler than the central regions and has $kT\lesssim 5$\,keV.

\subsection{Profiles across the edge}\label{profiles}

\subsubsection{Surface brightness profile}\label{SB_prof}

As a first step toward understanding the physical nature of the northern edge, we examine the surface brightness profile in that region with the aim of measuring the density jump at the edge. To that end we import the broken powerlaw density model \citep[described in detail in][]{owers2009c} into the {\it Sherpa} package and fit to the 0.5-7\,keV image near the edge. Initially, we fit to an annular sector centered at R.A.=15:10:11.93, decl. = 33:30:36.77 and with an angular range $50-130\degr$, where angles are measured north from west. The sector is shown as a white dashed region in the right panel of Figure~\ref{tmap}. The center is chosen as the approximate center of the radius of curvature of the edge, while the opening angle is chosen to match the range over which the edge is clearly visible. Given the complicated surface brightness distribution away from the edge and toward the cluster centre, and the limited accuracy of the powerlaw model approximation to the data at larger radii, we also constrain the radial range of the data fitted to $147\arcsec < r < 276\arcsec$. Because of the low counts per pixel, during fitting we minimize the CSTAT statistic and use a combination of Monte-Carlo and simplex methods for the minimization. We fit for the powerlaw slopes and amplitudes, the position of the edge and the ellipticity. The centroid of the model is fixed to the initial estimate of the center of curvature of the front, while the position angle is set to $90\degr$ to bisect the edge. A constant is added to the broken powerlaw model to account for the background. The best fitting values and their associated uncertainties are presented in Table~\ref{sb_fits}. The density jump across the edge is well approximated as the square root of the ratio of the surface brightness amplitudes just inside, $A_1$, and outside, $ A_2$, the edge, as determined during the fitting \citep{owers2009c}. This gives a density jump of $1.72^{+0.05}_{-0.04}$.

\begin{deluxetable*}{lccccccccc}
  \tablecaption{Parameters for the best fitting density models to the surface brightness across the northern edge.\label{sb_fits}}
  \tablecolumns{10}
  \tablehead{\colhead{Sector} & \colhead{Centroid} & \colhead{$\theta$ ($\Delta \theta$)} & \colhead{$R_{\rm f}$}& \colhead{Ellipticity} &\colhead{$\alpha_1$} & \colhead{$\alpha_2$}& \colhead{$A_1$} & \colhead{$A_2$} & \colhead{$\sqrt{A_1/A_2}$}\\
\colhead{} & \colhead{(J2000)} & \colhead{(degrees)} & \colhead{(kpc)}& \colhead{} &\colhead{} & \colhead{}& \colhead{($10^{-8}$)} & \colhead{($10^{-8}$)} & \colhead{}} \\
\startdata
Total & 15:10:11.93,  33:30:36.77 & 90 (50-130) & $402.0^{+0.4}_{-0.8}$ &$0.14^{+0.01}_{-0.01}$ &$1.50^{+0.10}_{-0.11}$ & $2.32^{+0.07}_{-0.07}$ & $7.30^{+0.22}_{-0.19}$& $2.46^{+0.12}_{-0.11}$ & $1.72^{+0.05}_{-0.04}$ \\
Nose & 15:10:11.93, 33:30:36.77 & 90 (76.67-103.33) & $404.4^{+0.9}_{-1.4}$ &$0.29^{+0.02}_{-0.04}$ &$1.14_{-0.15}^{+0.15}$ & $2.30_{-0.13}^{+0.13}$ & $8.02^{+0.34}_{-0.32}$& $2.40^{+0.21}_{-0.18}$ & $1.83^{+0.09}_{-0.08}$\\
East & 15:10:09.73,  33:30:17.07 & 116.67 (103.33-130) & $447.4^{+0.5}_{-1.7}$ &$0.00^{----}_{----}$ &$1.61_{-0.32}^{+0.23}$ & $2.57_{-0.16}^{+0.13}$ & $9.93^{+0.71}_{-0.45}$& $2.85^{+0.29}_{-0.20}$ & $1.86^{+0.12}_{-0.08}$\\
West & 15:10:11.93,  33:30:36.77 & 63.33 (50-76.67) & $391.6^{+1.1}_{-2.0}$ & $0.00^{----}_{----}$ & $1.32^{+0.23}_{-0.27}$ &$2.26_{-0.13}^{+0.14}$& $6.03^{+0.48}_{-0.35}$  & $2.17_{-0.18}^{+0.23}$ &$1.67^{+0.11}_{-0.08}$
\enddata
\tablecomments{$R_{\rm f}$ is the position of the edge. The East and West sectors have ellipticity fixed to zero. The $\alpha_1$ and $\alpha_2$ values are the powerlaw slopes inside and outside the edge, respectively. The $A_1$ and $A_2$ values give the surface brightness just inside and just outside the edge, respectively, and they have units ${\rm photons/cm^2/s/pixel}$ where one pixel is $1.968\arcsec \times 1.968\arcsec$. For a detailed description of the density model, see \citet{owers2009c}.}
\end{deluxetable*}

\begin{figure*}
\includegraphics[angle=0,width=0.5\textwidth]{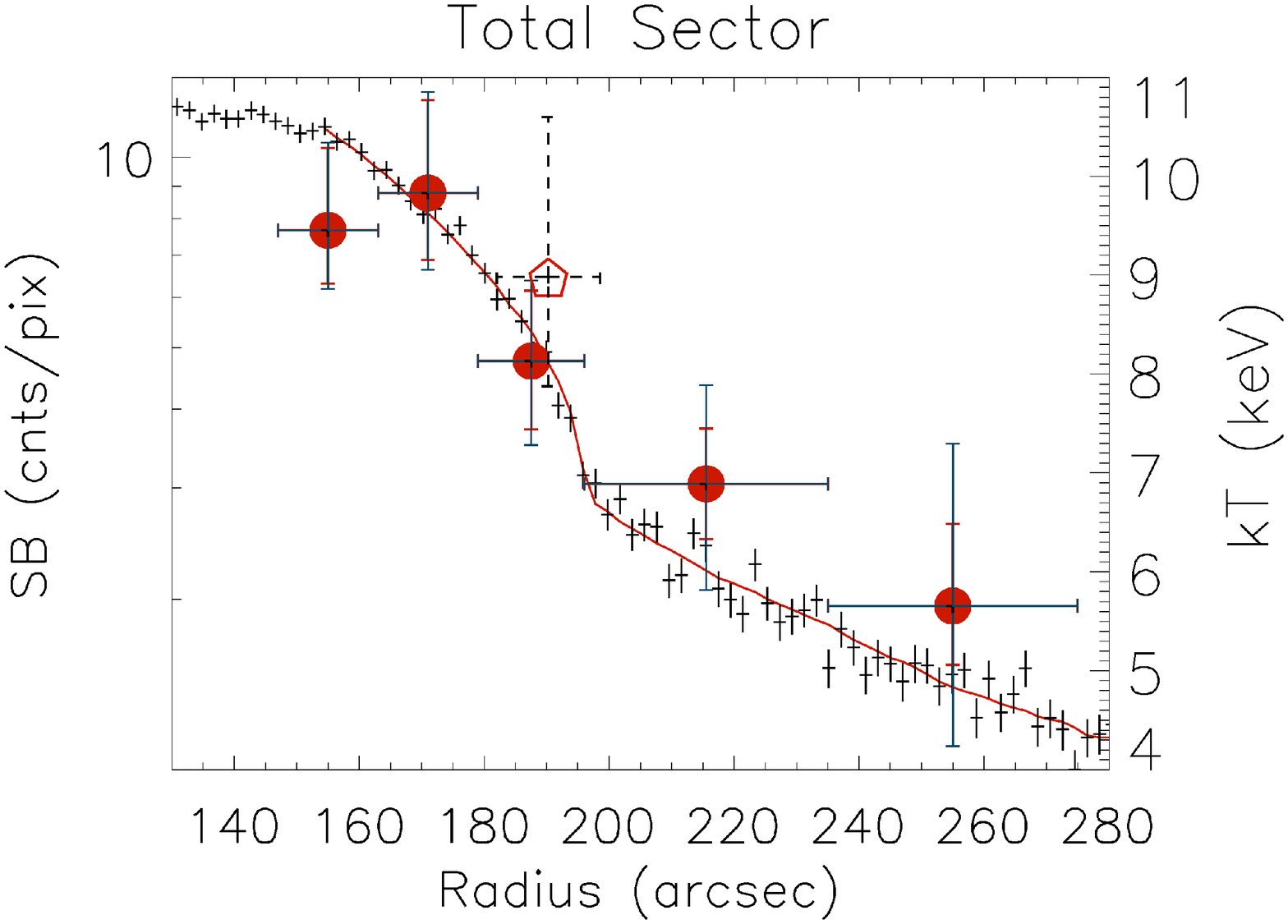}
\includegraphics[angle=0,width=0.5\textwidth]{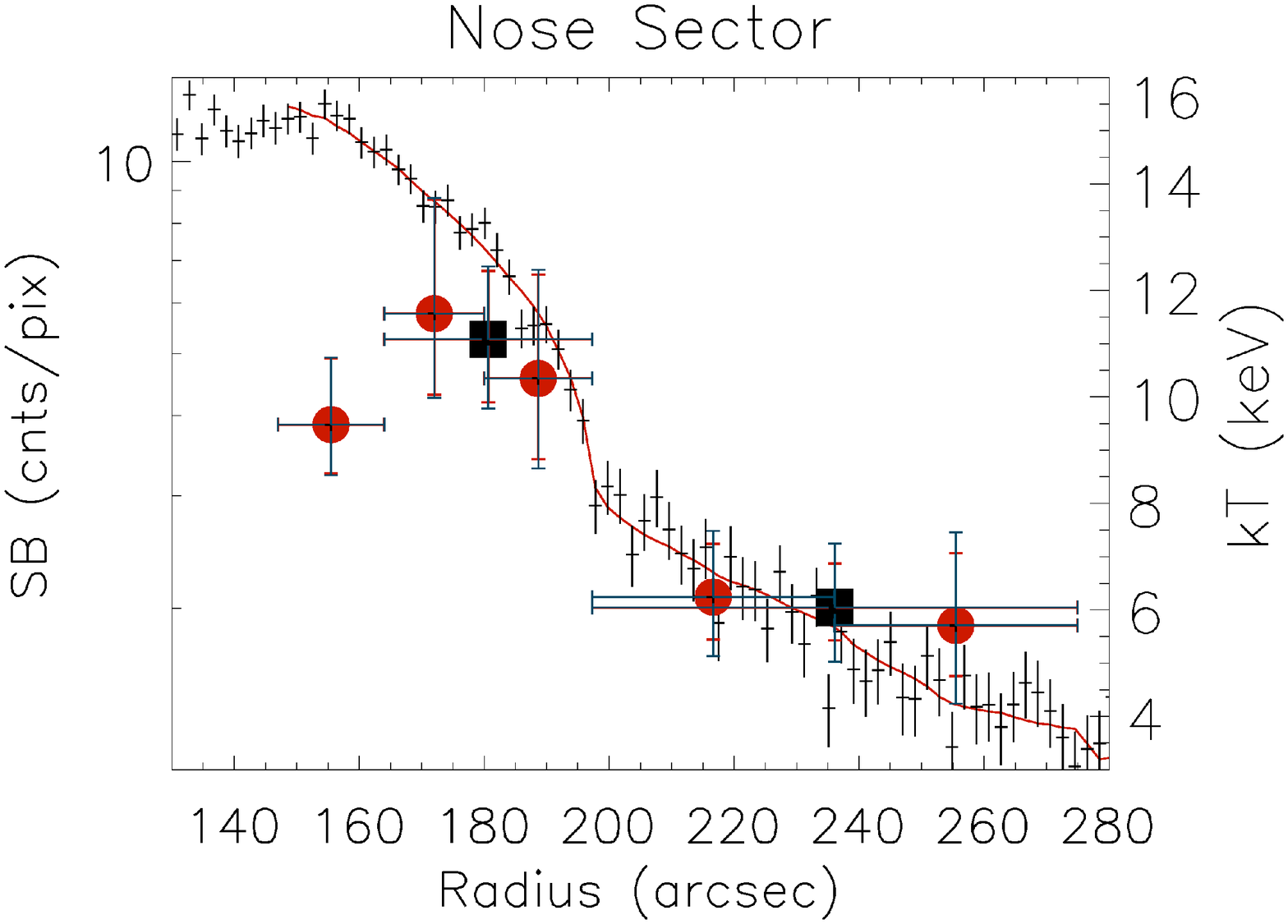}\\
\includegraphics[angle=0,width=0.5\textwidth]{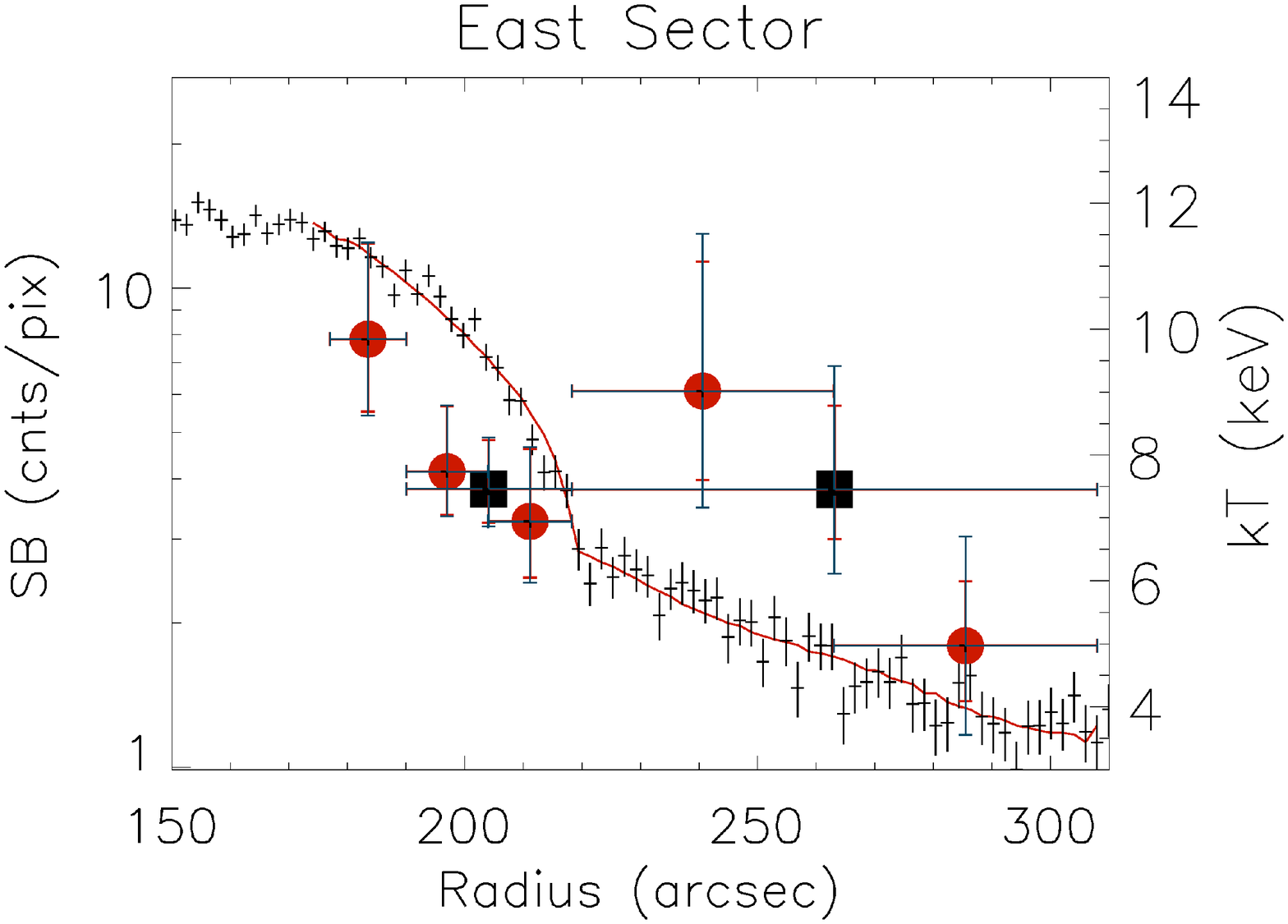}
\includegraphics[angle=0,width=0.5\textwidth]{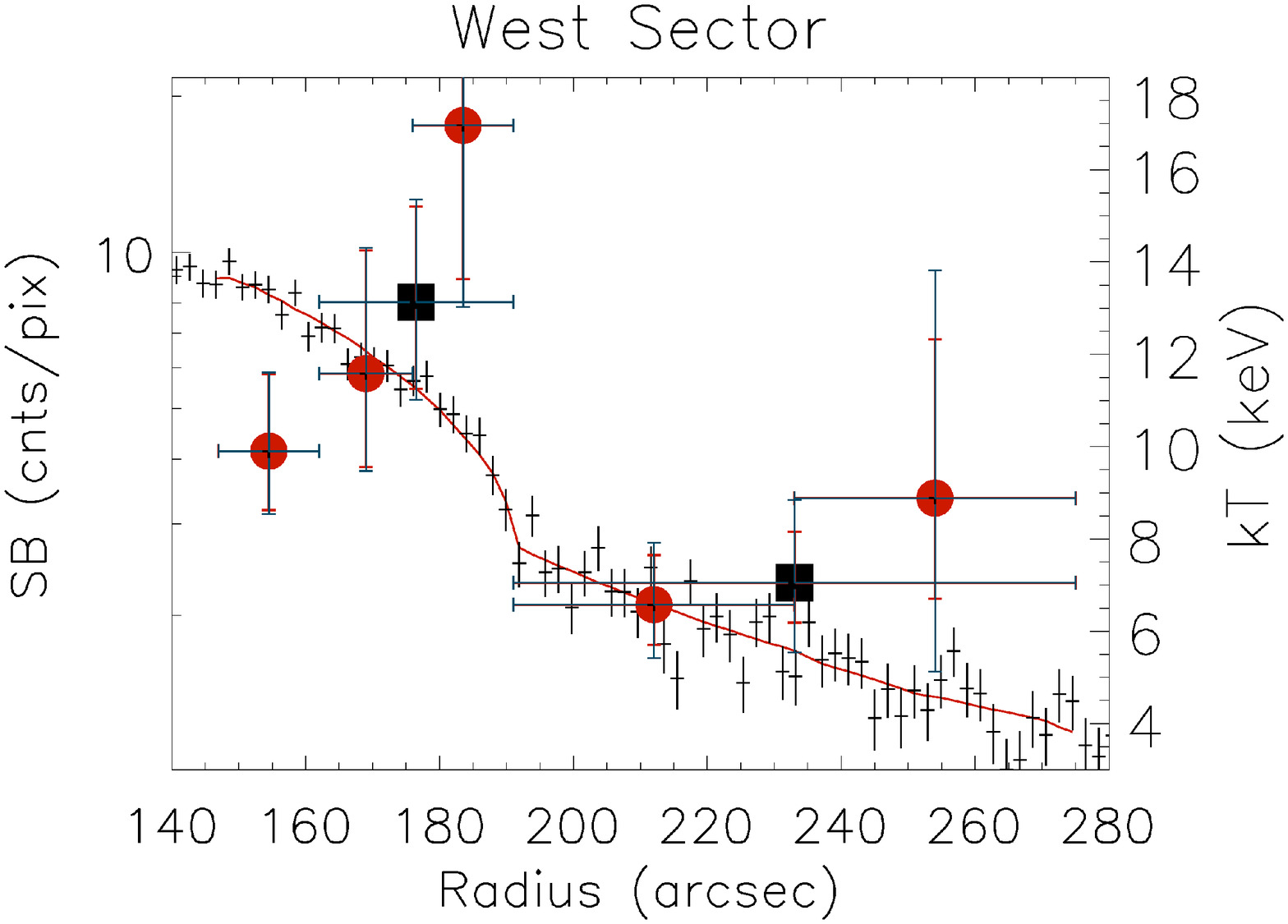}\\
\caption{In the four panels the black crosses show the surface brightness profiles across the edge for the total sector (top left panel) and divided across the three equi-angular sectors shown as a white dashed region in the right panel of Figure~\ref{tmap}: the nose sector (top right), the eastern sector (bottom left) and the western sector (bottom right).  The solid red line in each panel shows the best fitting density model. The parameters for the model can be found in Table~\ref{sb_fits}. The filled red circles show the temperature profile across the edge. The uncertainties are also shown, with the red ticks showing the $68\%$ statistical uncertainties, while the blue ticks show the impact of changing the background normalization by $\pm 10\%$. The black filled boxes show the best fitting temperatures from a simultaneous fit of the two spectra just inside and outside of the edge where the temperature is tied. In the top left panel, the open pentagon shows a deprojected temperature for the region just inside the edge. \label{SB_tot}}
\end{figure*}

\begin{figure}
\includegraphics[angle=0,width=.5\textwidth]{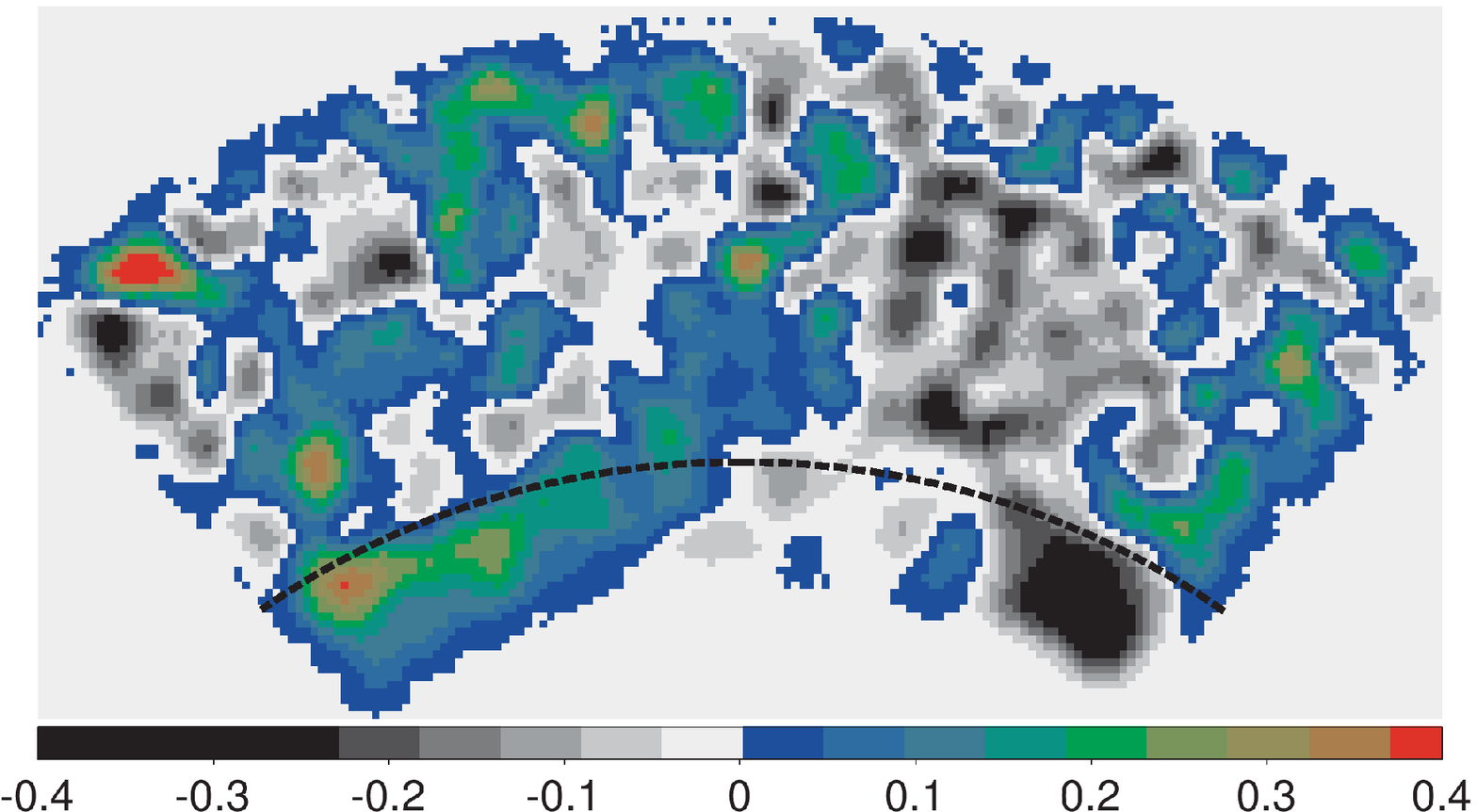}
\caption{Residual image generated by subtracting the best fitting density model from the data for the total sector shown in the top left panel of Figure~\ref{SB_tot}.  While the 1D fit is acceptable, there are clear azimuthal variations in surface brightness on the inside of the edge.\label{SB_tot_resid}}
\end{figure}

The radial surface brightness profile and the corresponding best fitting model are shown in Figure~\ref{SB_tot}. While the azimuthally collapsed 1D radial profile fit appears to be very good, inspection of the 2D residuals in Figure~\ref{SB_tot_resid} reveals significant brightness variations inside the front. In particular, the eastern portion of the edge shows strong  positive residuals while the western portion shows negative residuals. To further investigate the azimuthal variations in surface brightness along the edge, we divide the sector into three equi-angular regions ($\Delta \theta=26.67\degr$, see the white dashed regions in the right panel of Figure~\ref{tmap}) and refit the surface brightness model. The profiles from the nose, east and west regions are shown in the top right, bottom left and bottom right panels of Figure~\ref{SB_tot}, respectively. The corresponding model fits are shown in red, with the fitted parameters presented in Table~\ref{sb_fits}. The surface brightness just inside the edge for the eastern part of the sector is higher than both the nose and western parts, as can be seen by comparing the $A_1$ values presented in Table~\ref{sb_fits}. The radius of curvature is also different in the eastern part of the edge -- a different centroid was required in order to obtain a better fit to the surface brightness profile.

\subsubsection{Temperature Profiles}

Accurately determining the thermal properties across the edge is crucial for understanding its nature. For each of the annular sectors with surface brightness profiles presented in Figure~\ref{SB_tot}, we produce corresponding temperature profiles. We use the centroids and best-fitting ellipticities presented in Table~\ref{sb_fits} to define the elliptical annular sectors from which spectra are extracted. To minimize the impact of contamination from the bright emission just inside the edge on the temperature measurement for the region just outside the edge, we divide the inner and outer spectral regions at the best-fitting edge radii in Table~\ref{sb_fits}. The temperatures are plotted as solid red circles in Figure~\ref{SB_tot}, with the temperature scale presented on the right hand axis of each plot. The statistical $68$th percentile uncertainties are overplotted in red, while the blue error bars show the quadrature sum of the statistical uncertainties and the systematic temperature shift incurred by changing the background normalisation by $\pm 10\%$. The temperatures measured in the regions with the lowest surface brightness, i.e., the regions at larger radii, are most affected by the systematics induced by the change in the background normalization. The uncertainties on the temperature measurements for the regions inside the edge are dominated by the statistical uncertainty. For the ``nose'', ``east'' and ``west'' sectors, the filled boxes shown in Figure~\ref{sb_fits} are temperature measurements derived from a simultaneous fit to the spectra extracted from the two annuli immediately inside and immediately outside the edge. During the simultaneous fitting, the temperatures are tied and the normalizations are allowed to vary. The temperatures measured in each sector for the annuli immediately inside and outside the edge are presented in Table~\ref{temps}, as are the temperatures derived from the simultaneous fits.

The temperature profile for the ``Total'' sector reveals $\sim 9-10$\,keV gas in the first two annuli, dropping to $\sim 8$\,keV just inside the edge. The temperatures measured in the two annuli outside the edge are consistent within their uncertainties at $kT\sim 6-7$\,keV. While the best-fitting temperature is higher on the brighter side of the edge when compared with the fainter side, the difference is marginal and the two temperatures are consistent within their uncertainties. However, the temperature measurement just inside the edge is affected by the projection of cooler gas lying along the line of sight. To demonstrate the impact of this projected gas on the temperature measurement, we estimate the deprojected temperature just inside the edge, shown as an open pentagon in Figure~\ref{SB_tot}. The deprojection is performed in the manner described in \citet{owers2009c}. Briefly, a second MEKAL component is included in the fit to the spectra for the region just inside the edge. The temperature of this component is fixed at the best fitting temperature determined from the fit to the annular region just outside the edge. The normalisation  of this second component is also fixed to the best fitting normalisation determined in the region just outside the edge, but it is corrected by a scaling factor which accounts for the different emission measures expected from the different volumes probed. This scaling factor is determined from the best fitting density model. The deprojected temperature is $kT=9.0^{+1.6}_{-1.1}\,$keV and is consistent within the uncertainties with the projected measure of $kT=8.1^{+0.8}_{-0.9}\,$keV.

Considering the temperature profiles obtained in the equi-angular sectors, the ``nose'' and ``west'' sectors  (top right and bottom right panels in Figure~\ref{SB_tot}, respectively) show abrupt drops in temperature going from $\sim 12$\,keV on the inside to $\sim 6.5$\,keV on the outside of the edge. The differences in temperature are significant at the $> 2\sigma$ level (including the uncertainty due to the background normalization). Conversely, the temperature profile for the ``east'' sector shows an increase in temperature going from the inside to the outside of the edge, although the uncertainties on the outer measurement are large. However, the simultaneous fits to the larger radial ranges in the ``east'' sector indicate that the temperatures are consistent at 7.5\,keV across the edge.

\subsubsection{The nature of the edge}\label{edgenature}

Having determined the temperature and density on either side of the edge, we are now in a position to determine the physical nature of the edge as a shock or cold front. In the case of a cold front, the pressure is expected to be approximately continuous across the edge, a fact borne out in observations \citep{markevitch2000,owers2009c}. Thus, for an ideal gas where the pressure~$\propto n_{\rm e} kT$ the condition
\begin{equation}
{{\ktout} \over {\ktin}} \simeq {\rho_{\rm jump}}
\end{equation}
must hold across the edge where ${\rho}_{\rm jump}=\nein/\neout = \sqrt{A_1/A_2}$. Given the temperature measured on the inside of the edge, $\ktin$ (Table~\ref{temps}), we can use this condition to predict the temperature we expect to measure outside the edge, $\ktout$, assuming that the edge is a cold front. For each sector, we list the predicted $\ktout$ for a cold front in Table~\ref{temps}. In all but the ``east'' sector, the measured value for $\ktout$ is significantly lower than that predicted under the assumption of a cold front. This is also demonstrated by directly measuring the pressure jump at the edge, also listed in Table~\ref{temps}, where in the majority of cases the pressure jump is inconsistent with unity, with the only exception being the ``east'' sector which is marginally consistent with unity. We conclude that the edge cannot be a cold front.

Along similar lines, we can compare the measured properties across the edge under the assumption that it is caused by a shock front. In this case, a pressure jump is expected \citep[][]{markevitch2002} and the temperatures inside and outside the edge are related to the density jump by the Rankine-Hugoniot jump condition
\begin{equation}\label{ktjump}
{{\ktout} \over {\ktin}} = { {( \gamma + 1) - {\rho}_{\rm jump}( \gamma - 1)} \over ( \gamma + 1)-( \gamma - 1) {\rho}_{\rm jump}^{-1} }
\end{equation}
where $\gamma=5/3$ is the adiabatic index for a monatomic gas. Comparing the $\ktout$ expected in the case of a shock front with the observed values in Table~\ref{temps}, there is better agreement when compared with the cold front model for the ``total'', ``nose'' and ``west'' sectors. Again, there is a discrepancy between the shock model and the observations for the ``east'' sector. We note that for this part of the front, temperatures for the preshock spectra are affected by the hot region east of the front and those for the postshock spectra are affected by the cool filament discussed in Section~\ref{tempmaps}, neither of which appear to be related to the front. While it is difficult to understand the properties for the ``east'' sector under the assumption of these two models, it is clear that the remainder of the edge is consistent with being due to a shock front. In particular, under the shock interpretation it is within the ``nose'' sector that the observed and predicted density, temperature and pressure jumps should have best agreement. This is because it is within this sector that the assumptions used in determining the Rankine-Hugoniot relations, i.e. a one-dimensional shock with all of the gas motion being normal to the edge surface, are most likely to hold. Indeed, it is here that we see the best agreement between the observed properties and those expected if the edge is a shock. Therefore, we conclude that the edge is most likely a shock front.

\begin{deluxetable*}{lcccccc}
  \tablecaption{Temperatures and thermodynamic properties across the edge in sectors corresponding to those listed in Table~\ref{sb_fits}.\label{temps}}
  \tablecolumns{7}
  \tablehead{\colhead{Sector} & \colhead{$\ktin$} & \colhead{$\ktout$} &\colhead{$\ktout$}&\colhead{$\ktout$}&\colhead{Pressure jump}&\colhead{Counts}\\
    \colhead{} & & \colhead{Measured}& \colhead{Shock} & \colhead{Cold front} & (in/out)\\ 
}

\startdata
Total & $8.1_{-0.7/0.9}^{+0.7/0.8}$& $6.9_{-0.6/1.1}^{+0.6/1.0}$ &$5.4^{+0.5}_{-0.7}$ & $14.0^{+1.4}_{-1.6}$ & $2.0^{+0.3}_{-0.3}$ & 4752/4434\\
Nose & $10.4_{-1.5/1.7}^{+2.0/2.0}$ &$ 6.2_{-0.8/1.1}^{+1.0/1.2}$  & $6.5^{+1.3}_{-1.1}$& $19.0^{+3.8}_{-3.2}$& $3.1^{+0.9}_{-0.8}$& 1833/1458\\

Nose (1+2)& $ 11.1_{-1.2/1.3}^{+1.3/1.4}$ &$ 6.0_{-0.6/1.0}^{+0.8/1.2}$ & $7.0^{+0.9}_{-0.9}$& $20.3^{+2.8}_{-2.5}$& $3.4^{+0.8}_{-0.7}$& 1833+2571/1458+926\\

East & $ 6.9_{-0.9/1.0}^{+1.2/1.2}$& $ 9.0_{-1.4/1.9}^{+2.1/2.5}$ & $4.3^{+0.7}_{-0.9}$& $12.8^{+2.4}_{-1.9}$& $1.4_{-0.4}^{+0.5}$& 1390/1714\\

East (1+2)& $ 7.5_{-0.5/0.6}^{+0.8/0.8}$& $ 7.5_{-0.8/1.3}^{+1.3/2.0}$ & $4.6^{+0.5}_{-0.7}$& $14.0^{+1.7}_{-1.3}$& $1.9_{-0.4}^{+0.6}$& 1390+2395/1714+947\\

West & $ 17.0_{3.3/3.9}^{+5.5/5.6}$& $6.6_{-0.9/1.2}^{+1.1/1.4}$  & $11.6^{+3.8}_{-2.0}$& $28.4^{+9.5}_{-6.7}$& $4.3^{+1.7}_{-1.3}$& 1216/1525\\

West (1+2)& $ 13.1_{-1.9/2.1}^{2.1/2.2}$& $7.1_{-0.9/1.5}^{+1.1/1.8}$  & $9.0^{+1.5}_{-1.2}$& $21.9^{+4.0}_{-3.7}$& $3.1_{-0.8}^{+1.0}$& 1216+1540/1525+974

\enddata
\end{deluxetable*}

\subsection{Cluster membership and the spatial distribution of cluster members}\label{membership}

Previous studies have used photometrically defined cluster member samples to study the spatial distribution of galaxies in A2034, revealing that the distribution is bimodal \citep{okabe2008,vanweeren2011}. One of the local peaks in galaxy surface density is associated with BCG1 in the cluster core.  A second local peak in the galaxy density is located just ahead of the northern edge and coincident with BCG2 (which we note is the third brightest member galaxy; magenta diamond point in the left panel of Figure~\ref{tmap}). \citet{okabe2008} also used weak lensing to construct projected mass maps which reveal that the central regions contain an irregular mass distribution with multiple mass concentrations. They find a significant mass structure coincident with the second local peak just ahead of the northern edge. Here, we determine the spatial distribution of galaxies in A2034 using a sample of spectroscopically confirmed cluster members.

\begin{figure}
\includegraphics[angle=0,width=0.45\textwidth]{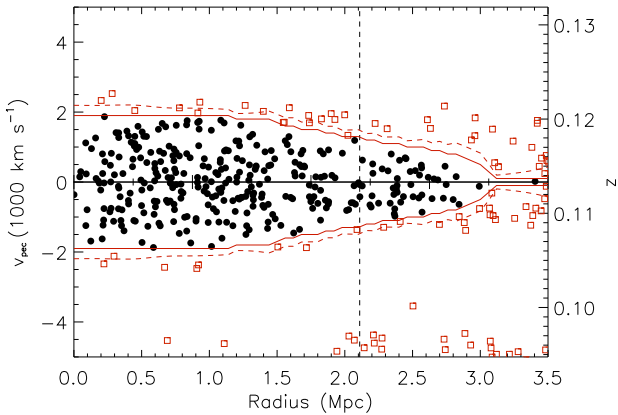}
\caption{Caustics membership allocation. Filled Black circles show cluster members, open red squares show non-members. The red lines show the caustics which define the member/non-member boundaries and the red dashed lines show the $1\sigma$ uncertainties. The radius is measured with respect to BCG1.\label{mem_sel}}
\end{figure}

To build a sample of spectroscopically confirmed cluster members, we combine our MMT Hectospec data described in Section~\ref{hecto_data} with 259 redshifts taken from SDSS DR9 \citep{ahn2012} and 81 from the Hectospec Cluster Redshift Survey \citep{rines2012}. Membership allocation is determined in redshift-radius phase-space using the caustics method as outlined in \citet{owers2013} and described in detail elsewhere \citep{diaferio1999,serra2013}. Figure~\ref{mem_sel} shows the peculiar velocity as a function of radius and the caustic boundaries used to define membership. Our sample of spectroscopically confirmed members contains 328 galaxies within 3.5\,Mpc of BCG1. Using this sample, we measure a cluster redshift of $z_{\rm clus}=0.1132 \pm 0.0002$ and a velocity dispersion of $\sigma_{\rm v} = 846 \pm 31\,$\kms. While the shape of the velocity distribution does not depart significantly from a Gaussian shape, the velocity dispersion measured here may be significantly affected by the merger occuring in A2034. Therefore, this dispersion may not reflect the virialized motion of galaxies as it does in more relaxed systems. From the caustics we also measure a mass within $r_{200}$ of $M_{200}=1.1 \pm 0.4 \times 10^{15}\,$\msolar, where $r_{200}\simeq\sqrt{3} \sigma_{\rm v}/10H(z) = 2.1\,$Mpc is the cluster radius where the mean density enclosed is 200 times the critical density of the Universe at the cluster redshift \citep{carlberg1997}.

\begin{figure}
\includegraphics[angle=0,width=0.47\textwidth]{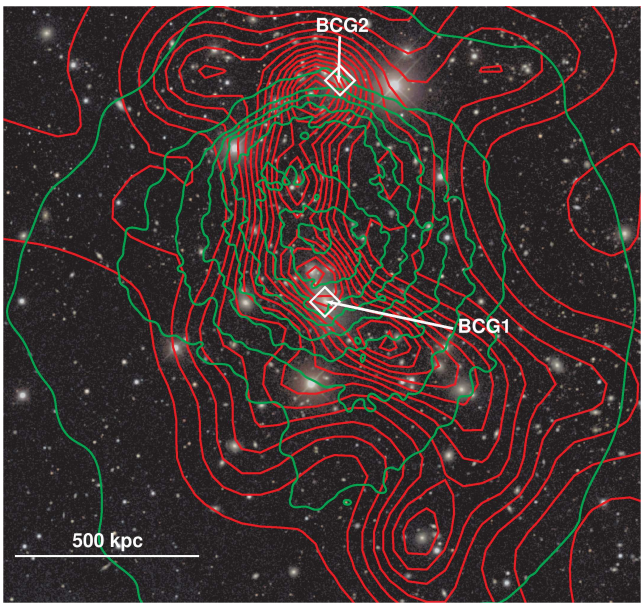}
\caption{SDSS RGB image with the red contours showing the adaptively smoothed surface member galaxy isopleths. Contour levels begin at 25 gal/Mpc$^2$ and increment in intervals of 10 gal/Mpc$^2$. The green contours show the same \chan\, surface brightness shown in Figure~\ref{opt_chan}, but with only half of the intervals plotted. \label{dense_chan}}
\end{figure}

The red contours overlaid onto an SDSS RGB image in Figure~\ref{dense_chan} show the adaptively smoothed surface density of spectroscopically confirmed cluster members. The adaptive smoothing is performed as outlined in \citet{owers2013} and only the galaxies within 1.5\,Mpc of BCG1 are included. Considering the central part of the cluster, we find a local peak in the galaxy surface density located within $\sim 75\,$kpc of BCG1. This galaxy overdensity is offset by 103\,kpc to the south of the peak in the X-ray surface brightness. The projected mass map of \citet{okabe2008} reveals an irregular distribution of mass with two significant peaks located within $\sim 170\,$kpc of, and likely associated with, BCG1. The most significant local peak in \citeauthor{okabe2008}'s mass map lies just south of the peak in the X-ray surface brightness, while the second local peak lies $\sim 140\,$kpc to the south of BCG1. 

Considering the region north of the cluster centre and near the shock,  we confirm that there is a local peak in the galaxy surface density located within $\sim 65\,$kpc of BCG2, which is located just ahead of the nose of the shock to the north. This local peak in the galaxy surface density also coincides with a significant local peak in the projected mass map in \citet{okabe2008}. This provides strong evidence for a significantly massive structure associated with BCG2.

Aside from the two prominent substructures associated with BCG1 and BCG2, there are two less significant local peaks in the galaxy surface density located in the X-ray bright part of the cluster. One is located at the midpoint between BCG1 and BCG2 and the other is located just to the southwest of BCG1. Neither of these structures are associated with bright cluster members, nor do they coincide with any of the lensing features. Therefore, we do not consider them to be associated with dynamically important substructures. Moreover, we note that we were not able to obtain a redshift for every target in our initial photometric sample. These less significant structures may be artifacts of the redshift incompleteness. However, we note that consistency of our results and those of \citet{okabe2008} and \citet{vanweeren2011} indicates that the substructures associated with BCG1 and BCG2 are real.

\section{Discussion}\label{discussion}

The deeper \chan\ data presented here reveal in more detail the merger-related features discussed in \citetalias{kempner2003}. The simplest interpretation of the data is that A2034 has undergone a two-body merger viewed close to the plane of the sky and just after a low impact parameter core passage. The merger axis is aligned along a north-south direction with the secondary subcluster currently traveling to the north and away from the primary cluster. Here we discuss the evidence supporting this scenario.

\subsection{Shock Mach number and geometry}\label{mach}
Perhaps the strongest evidence for the proposed merger scenario comes from the edge to the north. The temperature and density measurements across the edge (Section~\ref{profiles}) show that it cannot be a cold front and that its properties are consistent with it being a merger shock. This shock is presumably driven by the subcluster revealed as a local overdensity in the galaxy distribution. This provides support for supersonic motion to the north and we can now use the information provided by the shock jump to estimate the merger geometry and the time since pericenter. The Rankine-Hugoniot shock jump conditions can be used to determine the shock velocity where the Mach number
\begin{equation}\label{mach_eqn}
M = v_{\rm shock}/c_{\rm s} = {\left[ {2 {\rho}_{\rm  jump}} \over {( \gamma + 1) - {\rho}_{\rm jump} ( \gamma - 1)} \right]^{1/2}},
\end{equation}
in the notation of Section~\ref{edgenature}, where $v_{\rm shock}$ is the shock speed and the sound speed in the unshocked medium is  $c_{\rm s}=\sqrt{\gamma k_{\rm B} T_{\rm out}/\mu m_{\rm p}}$. Using $\rho_{jump}=1.83^{+0.09}_{-0.08}$ from the surface brightness fit to the ``nose'' sector in Section~\ref{SB_prof}, we find $M=1.59^{+0.06}_{-0.07}$. The sound speed in the pre-shock gas, where $\ktout = 6.0\,$keV, is $c_{\rm s} = 1293$\kms\ giving a shock velocity of $v_{\rm shock}=2057$\kms. We can check the Mach number measurement by inverting Equation~\ref{ktjump} to determine the expected density jump based on the temperature jump in the nose sector (Table~\ref{temps}). Inserting this expected density jump into Equation~\ref{mach_eqn} gives $M=1.82^{+0.37}_{-0.32}$, consistent within the uncertainties with the Mach number derived using the density jump.  The sharpness of the edge indicates that the shock motion is largely perpendicular to our line of sight (LOS). Assuming that BCG2, just north of the nose of the shock, is the central galaxy of the infalling subcluster, we can use it to estimate the component of shock motion along our LOS.  Its proximity to the shock suggests that it is moving at a similar velocity. The central BCG, BCG1, has $z=0.1118$ ($v_{\rm pec}=-362\,$\kms\ with respect to $z_{\rm clus}$) while BCG2 has $z=0.1151$ ($v_{\rm pec}=529\,$\kms). The difference in peculiar velocity is $\Delta v_{\rm pec} = 891\,$\kms\ which indicates that some of the motion of the shock may be directed along our LOS. Assuming that the peculiar velocities of these two galaxies are representative of their associated substructures, and that the shock and substructure galaxies are comoving, the velocities can be used to constrain the inclination to the plane of the sky $\theta_{\rm inc}=\arctan(891/2057) = 23\degr$. Given the shock velocity, the inclination angle and the distance from the  BCG1 ($\sim 588\,$kpc) we estimate that the core passage occurred $\sim 0.3$\,Gyr ago, similar to the timescales estimated for the Bullet and A2146 \citep[][]{markevitch2002,russell2010}. This evidence reinforces the hypothesis that the merger in A2034 is observed shortly after core passage, and is occurring on an axis which is close to the plane of the sky.

It is interesting to note that the surface brightness profiles across the shock presented in Section~\ref{SB_prof} indicate that the eastern portion of the shock is brighter than the western portion. The cause of this asymmetry is unclear, although we present two hypotheses that may explain this. The first hypothesis is that the distribution of the pre-shocked ICM is asymmetric about the nose of the shock. In this scenario, the pre-shocked ICM to the east of the nose is denser than that to the west. This will generally lead to a higher density in the postshock gas there. Since the X-ray surface brightness $\propto \int n^2_e dl$, where $l$ is the distance along the LOS, the surface brightness will be enhanced. An asymmetric distribution of pre-shocked ICM about the nose of the shock can occur if the merger is not directly head-on. An excellent example of this effect can be seen in the simulations of \citet[][e.g., their Figure 3]{mastropietro2008}. However, for A2034 the peak of the X-ray emission and the nose of the shock lie approximately on a line connecting BCG1 and BCG2. This indicates that the impact parameter must not have been large. Alternatively, if the ICM associated with A2034 was initially ellipsoidal with major axis running from southwest to northeast then this would produce the desired pre-shocked ICM asymmetry for a head-on merger. The second hypothesis is that the shock may be propagating through an inhomogenous and/or non-static ICM, as suggested by \citet{mazzotta2011} to explain the ``M-shaped'' shock in the cluster RXJ1314.4-2512. This may distort the shape of the shock front, leading to an asymmetric brightness distribution.

\subsection{Galaxy distribution}
The spatial distribution of the galaxies in comparison to the ICM also supports our favored merger scenario. During the core passage phase of a head on merger, the ICM of the subclusters can be ram-pressure stripped from the collisionless galaxies and dark matter. This leads to an offset in the X-ray and galaxy centroid which is a strong indicator of the direction of motion. The most prominent example of this is seen in the Bullet cluster \citep{barrena2002,clowe2006}, although counter-examples exist \citep[][]{hallman2004, russell2010, owers2011a}. Considering the central cluster region, Figure~\ref{dense_chan} shows that BCG1 and the peak in galaxy surface density are approximately cospatial, while the main peak in the X-ray emission located 172\,kpc away just east of due north. In our scenario, this offset occurred when the secondary subcluster associated with BCG2 traversed the core from the south on its way to the observed position in the north. The force on the gas due to this high speed encounter has displaced the ICM from the core of the main cluster and BCG1. However, the separation of the gas and dark matter components does not appear to be as clean as that seen in the Bullet cluster. The projected distribution of mass seen in the lensing maps of \citet{okabe2008} is complex with two significant mass clumps in the core, neither of which is spatially coincident with BCG1. Nevertheless, the clumpy mass distribution indicates a strongly disturbed structure indicative of recent merger activity.

Similarly, the gas and galaxies associated with the secondary substructure were also separated, with the secondary's galaxies now coincident with the nose of the shock cone. Unlike in the Bullet cluster, we do not observe a  prominent dense cool core trailing the secondary's galaxies. If the gas core did not harbor a dense, cool core prior to the merger, \citep[as is the case for around $25-30\%$ of local clusters; see ][]{hudson2010} then it would have been more prone to disruption during the core passage. There does appear to be clumpy emission located just south of the shock which is cooler than the surrounding shock heated gas. We speculate that this is the remnant of the gas core previously associated with the secondary and that the core has been completely disrupted by the merger.

Further evidence for the removal of gas from the secondary may come from the tail of excess emission located south of BCG1 (the south excess in Figure~\ref{chan_img}). \citetalias{kempner2003}'s favored interpretation is that this emission comes from a background cluster. Alternatively, \citetalias{kempner2003} suggest that the southern excess may be associated with another subcluster merger in A2034. We do see a local peak in the galaxy surface density $\sim 685\,$kpc south of BCG1 which coincides with a $r=16.8$ magnitude galaxy and may be associated with the southern excess. However, in our merger scenario this emission can be interpreted as gas which has been stripped from the outer parts of the BCG2 subcluster as it fell from the south toward the primary core. 

In N-body hydrodynamic models for the Bullet Cluster \citep{springel2007, mastropietro2008}, the shock runs ahead of the subcluster during the initial infall phase of the merger. After core passage, when the subcluster's gas core is displaced from the galaxies and dark matter, the remnant gas core of the infalling subcluster acts as the ``piston'' driving the shock. If the gas core is disrupted as we suggest for A2034, the shock can weaken and slow, allowing the galaxies and dark matter to overtake it. Thus, we speculate that the merger shock in A2034 may be dying, consistent with its relatively low Mach number \citep[compared with, e.g., the M=2-3 shocks in other clusters;][]{markevitch2010} and the position of BCG2 ahead of the front. The simulations have also shown that the ICM of the main cluster is accelerated towards the infalling subcluster, boosting the speed of the shock in the gas significantly compared to its speed through the cluster \citep{springel2007}.  Thus, the speed through the cluster of the remnant galaxies may be significantly less than the shock speed of $2057\rm\ km\ s^{-1}$.  If so, we may also have underestimated the time since core passage of the merger. We also caution that a lower shock speed, and any relative motion between the shock and the galaxies ahead of the shock, will affect the inclination angle estimated in Section~\ref{mach}.

\subsection{Radio emission}

\begin{figure}
\includegraphics[angle=0,width=0.47\textwidth]{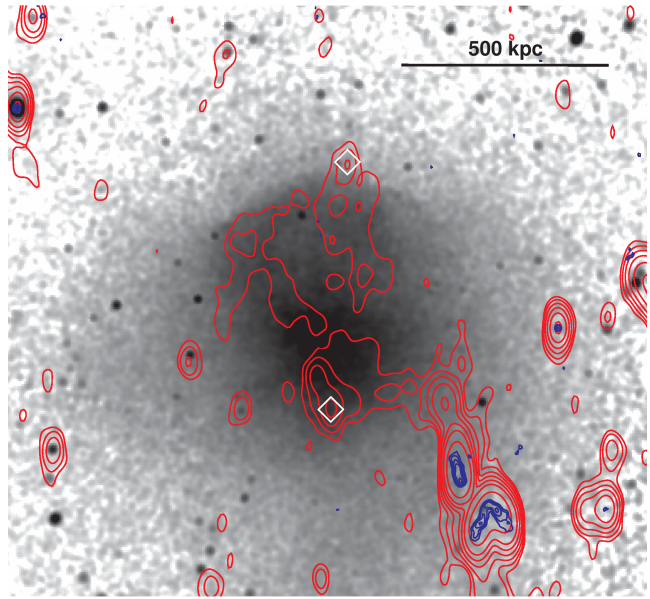}
\caption{\chan\ image (as in Figure~\ref{chan_image}) with 1.4\,GHz radio contours overlaid. The red contours show the WRST data presented in \citet{vanweeren2011}. These data are sensitive to the diffuse, low surface brightness emission and the contours have levels $24\,\mu~{\rm Jy~beam^{-1}}\times [4, 8, 32, 64, 128, 356]$. The blue contours show the higher resolution VLA-FIRST data. The contour levels are $0.15\,{\rm mJy beam^{-1}}\times [4, 8, 32, 64, 128, 356]$. The white diamond points show the positions of the BCGs as in Figure~\ref{opt_chan}. Note the diffuse emission located near the shock edge in the north revealed by the WRST data, and the head-tail and wide-angle-tailed radio galaxies revealed by the FIRST data. \label{radio}}
\end{figure}

Previous studies have shown that there is diffuse radio emission in the vicinity of the shock to the north \citep{kempner2001,rudnick2009,giovannini2009,vanweeren2011}. The classification of this diffuse radio emission is unclear. Both \citet{kempner2001} and \citet{vanweeren2011} suggest that the diffuse emission may be a radio relic, although \citet{rudnick2009} classify it as a radio halo. In Figure~\ref{radio} we overlay two sets of 1.4GHz radio contours onto a \chan\ image. The red contours are the same Westerbork Synthesis Radio Telescope (WRST) data presented in \citet{vanweeren2011}, while the blue contours are taken from VLA-FIRST data \citep{becker1995}. The VLA-FIRST data show a head-tail galaxy and wide-angle-tailed radio galaxy located to the southwest of the cluster core. Both of these galaxies are cluster members and their radio properties suggest high relative motion with respect to the ICM. The deeper WRST data reveal that the diffuse emission has an irregular morphology, while there is point source emission associated with both BCG1 and BCG2. In the region near the shock, there is weak, diffuse radio emission just behind the nose and eastern part of the front.

Given the irregular radio morphology \citep[compare with that seen in][]{vanweeren2010} and the weak spatial correlation with the shock edge, we suggest that the emission is due to the revival of pre-existing radio plasma which has been overrun by the shock, rather than being a consequence of direct acceleration of thermal electrons at the shock front. The direct acceleration scenario requires Mach numbers $M > 3$ to produce observable radio emission \citep{hoeft2007,kang2012}. Therefore, the low Mach number measured in Section~\ref{mach} provides further evidence against the direct acceleration scenario in A2034 \citep[as in A754;][]{marcario2011}. The fossil radio plasma can be revived by a shock through either the reacceleration of the mildly relativistic electrons \citep{markevitch2005}, or due to adiabatic compression of the plasma \citep{ensslin2001}. The reacceleration scenario predicts a single powerlaw spectrum with a slope related to the Mach number of the shock \citep{markevitch2005}, whereas the compression scenario predicts a much steeper slope at higher frequencies. Therefore, deep multi-band radio observations of A2034, which can be used to measure the integrated and spatially resolved spectral indices, will  help to constrain the nature of the emission and its relation to the shock \citep[e.g., as in ][]{giacintucci2008,vanweeren2010,vanweeren2011a,clarke2013}.  

On the classification of the source as a halo or relic, the passage of a shock which excites a radio relic should also align the magnetic field so that it is perpendicular to the shock surface \citep{ensslin1998}. This leads to ordered polarization vectors \citep[as observed in CIZAJ2242.8+5301;][]{vanweeren2010}, whereas in radio halos the polarization is much weaker and has less well-ordered polarization vectors \citep{feretti2012}. As pointed out in \citet{vanweeren2011}, polarimetric observations will have the potential to distinguish radio halo emission from shock-related relic emission in A2034.

\section{Conclusion}\label{conclusion}
We present a deep 250\,ks exposure of the merging cluster A2034. A previous study using shallower \chan\ data \citepalias{kempner2003} revealed a sharp edge to the north of the X-ray peak which was interpreted as being due to a cold front. Our deeper data have revealed that this interpretation cannot be correct, since the pressure is not continuous across any part of the edge. We show that the properties across the edge are consistent with it being caused by a shock with Mach number $M=1.59^{+0.06}_{-0.07}$, corresponding to a shock velocity of $\sim 2057$\kms. We supplement our \chan\ data with 328 spectroscopically confirmed cluster members drawn mainly from our MMT/Hectospec survey of the cluster. The spatial distribution of cluster members shows a complex distribution of local peaks in galaxy surface density. Notably, we confirm the existence of a substructure located at the nose of the shock cone which is the remnant of the merging substructure responsible for the shock. These observations indicate that A2034 is a cluster merger with a merger axis viewed within $\sim 23\degr$ of the plane of the sky. The subclusters are moving apart along a north-south direction approximately $0.3\,$Gyr after a low impact parameter core passage. The subcluster's gas core, which acted as the piston driving the shock, appears to have been completely destroyed during the core passage. The disruption of the piston may have led to a weakening of the shock which we speculate is now dying. Given its proximity, the possible radio relic is likely a result of the shock, although deeper radio observations are required to confirm its nature.

\acknowledgements
The authors thank Maxim Markevitch and Trevor Ponman for helpful discussions. We thank the referee, Herv\'e Bourdin, for his constructive suggestions. M.S.O. acknowledges the funding support from the Australian Research Council through a Super Science Fellowship (ARC FS110200023). Support for this work was provided by the National Aeronautics and Space Administration through Chandra Award Number GO1-12161X issued by the Chandra X-ray Observatory Center, which is operated by the Smithsonian Astrophysical Observatory for and on behalf of the National Aeronautics Space Administration under contract NAS8-03060. P.E.J.N. was partly supported by NASA contract NAS8-03060. W.J.C. acknowledges the funding support of the Australian Research Council through an Australian Professorial Fellowship throughout most of this work. R.J.vW is supported by NASA through the Einstein Postdoctoral grant number PF2-130104 awarded by the Chandra X-ray Center, which is operated by the Smithsonian Astrophysical Observatory for NASA under contract NAS8-03060. This research has made use of software provided by the Chandra X-ray Center (CXC) in the application packages CIAO, ChIPS, and Sherpa and also of data obtained from the Chandra archive at the NASA Chandra X-ray Center (cxc.harvard.edu/cda/). Some of the observations reported here were obtained at the MMT Observatory, a joint facility of the Smithsonian Institution and the University of Arizona. We thank the MMT operators and queue-schedule mode scientists for their help during observations and the staff at the Harvard-Smithsonian Center for Astrophysics Telescope Data Center for reducing the Hectospec data.

{\it Facilities:} \facility{CXO (ACIS)}, \facility{MMT (Hectospec)}


\begin{thebibliography}{65}
\expandafter\ifx\csname natexlab\endcsname\relax\def\natexlab#1{#1}\fi

\bibitem[{{Abazajian} {et~al.}(2009){Abazajian}, {Adelman-McCarthy},
  {Ag{\"u}eros}, {Allam}, {Allende Prieto}, {An}, {Anderson}, {Anderson},
  {Annis}, {Bahcall}, \& et~al.}]{abazajian2009}
{Abazajian}, K.~N., {Adelman-McCarthy}, J.~K., {Ag{\"u}eros}, M.~A., {Allam},
  S.~S., {Allende Prieto}, C., {An}, D., {Anderson}, K.~S.~J., {Anderson},
  S.~F., {Annis}, J., {Bahcall}, N.~A., \& et~al. 2009, \apjs, 182, 543

\bibitem[{{Ahn} {et~al.}(2012){Ahn}, {Alexandroff}, {Allende Prieto},
  {Anderson}, {Anderton}, {Andrews}, {Aubourg}, {Bailey}, {Balbinot}, {Barnes},
  \& et~al.}]{ahn2012}
{Ahn}, C.~P., {Alexandroff}, R., {Allende Prieto}, C., {Anderson}, S.~F.,
  {Anderton}, T., {Andrews}, B.~H., {Aubourg}, {\'E}., {Bailey}, S.,
  {Balbinot}, E., {Barnes}, R., \& et~al. 2012, \apjs, 203, 21

\bibitem[{{Anders} \& {Grevesse}(1989)}]{anders1989}
{Anders}, E., \& {Grevesse}, N. 1989, \gca, 53, 197

\bibitem[{{Arnaud}(1996)}]{arnaud1996}
{Arnaud}, K.~A. 1996, in Astronomical Society of the Pacific Conference Series,
  Vol. 101, Astronomical Data Analysis Software and Systems V, ed.
  {G.~H.~Jacoby \& J.~Barnes}, 17--+

\bibitem[{{Barrena} {et~al.}(2002){Barrena}, {Biviano}, {Ramella}, {Falco}, \&
  {Seitz}}]{barrena2002}
{Barrena}, R., {Biviano}, A., {Ramella}, M., {Falco}, E.~E., \& {Seitz}, S.
  2002, \aap, 386, 816

\bibitem[{{Becker} {et~al.}(1995){Becker}, {White}, \& {Helfand}}]{becker1995}
{Becker}, R.~H., {White}, R.~L., \& {Helfand}, D.~J. 1995, \apj, 450, 559

\bibitem[{{Brunetti} {et~al.}(2008){Brunetti}, {Giacintucci}, {Cassano},
  {Lane}, {Dallacasa}, {Venturi}, {Kassim}, {Setti}, {Cotton}, \&
  {Markevitch}}]{brunetti2008}
{Brunetti}, G., {Giacintucci}, S., {Cassano}, R., {Lane}, W., {Dallacasa}, D.,
  {Venturi}, T., {Kassim}, N.~E., {Setti}, G., {Cotton}, W.~D., \&
  {Markevitch}, M. 2008, \nat, 455, 944

\bibitem[{{Carlberg} {et~al.}(1997){Carlberg}, {Yee}, \&
  {Ellingson}}]{carlberg1997}
{Carlberg}, R.~G., {Yee}, H.~K.~C., \& {Ellingson}, E. 1997, \apj, 478, 462

\bibitem[{{Clarke} {et~al.}(2013){Clarke}, {Randall}, {Sarazin}, {Blanton}, \&
  {Giacintucci}}]{clarke2013}
{Clarke}, T.~E., {Randall}, S.~W., {Sarazin}, C.~L., {Blanton}, E.~L., \&
  {Giacintucci}, S. 2013, \apj, 772, 84

\bibitem[{{Clowe} {et~al.}(2006){Clowe}, {Brada{\v c}}, {Gonzalez},
  {Markevitch}, {Randall}, {Jones}, \& {Zaritsky}}]{clowe2006}
{Clowe}, D., {Brada{\v c}}, M., {Gonzalez}, A.~H., {Markevitch}, M., {Randall},
  S.~W., {Jones}, C., \& {Zaritsky}, D. 2006, \apjl, 648, L109

\bibitem[{{Diaferio}(1999)}]{diaferio1999}
{Diaferio}, A. 1999, \mnras, 309, 610

\bibitem[{{Dickey} \& {Lockman}(1990)}]{dickey1990}
{Dickey}, J.~M., \& {Lockman}, F.~J. 1990, \araa, 28, 215

\bibitem[{{Donnelly} {et~al.}(1998){Donnelly}, {Markevitch}, {Forman}, {Jones},
  {David}, {Churazov}, \& {Gilfanov}}]{donelly1998}
{Donnelly}, R.~H., {Markevitch}, M., {Forman}, W., {Jones}, C., {David}, L.~P.,
  {Churazov}, E., \& {Gilfanov}, M. 1998, \apj, 500, 138

\bibitem[{{En{\ss}lin} {et~al.}(1998){En{\ss}lin}, {Biermann}, {Klein}, \&
  {Kohle}}]{ensslin1998}
{En{\ss}lin}, T.~A., {Biermann}, P.~L., {Klein}, U., \& {Kohle}, S. 1998, \aap,
  332, 395

\bibitem[{{En{\ss}lin} \& {Gopal-Krishna}(2001)}]{ensslin2001}
{En{\ss}lin}, T.~A., \& {Gopal-Krishna}. 2001, \aap, 366, 26

\bibitem[{{Ettori} \& {Fabian}(2000)}]{ettori2000}
{Ettori}, S., \& {Fabian}, A.~C. 2000, \mnras, 317, L57

\bibitem[{{Fabricant} {et~al.}(2005){Fabricant}, {Fata}, {Roll}, {Hertz},
  {Caldwell}, {Gauron}, {Geary}, {McLeod}, {Szentgyorgyi}, {Zajac}, {Kurtz},
  {Barberis}, {Bergner}, {Brown}, {Conroy}, {Eng}, {Geller}, {Goddard},
  {Honsa}, {Mueller}, {Mink}, {Ordway}, {Tokarz}, {Woods}, {Wyatt}, {Epps}, \&
  {Dell'Antonio}}]{fabricant2005}
{Fabricant}, D., {Fata}, R., {Roll}, J., {Hertz}, E., {Caldwell}, N., {Gauron},
  T., {Geary}, J., {McLeod}, B., {Szentgyorgyi}, A., {Zajac}, J., {Kurtz}, M.,
  {Barberis}, J., {Bergner}, H., {Brown}, W., {Conroy}, M., {Eng}, R.,
  {Geller}, M., {Goddard}, R., {Honsa}, M., {Mueller}, M., {Mink}, D.,
  {Ordway}, M., {Tokarz}, S., {Woods}, D., {Wyatt}, W., {Epps}, H., \&
  {Dell'Antonio}, I. 2005, \pasp, 117, 1411

\bibitem[{{Feretti} {et~al.}(2012){Feretti}, {Giovannini}, {Govoni}, \&
  {Murgia}}]{feretti2012}
{Feretti}, L., {Giovannini}, G., {Govoni}, F., \& {Murgia}, M. 2012, \aapr, 20,
  54

\bibitem[{{Fruscione} {et~al.}(2006){Fruscione}, {McDowell}, {Allen},
  {Brickhouse}, {Burke}, {Davis}, {Durham}, {Elvis}, {Galle}, {Harris},
  {Huenemoerder}, {Houck}, {Ishibashi}, {Karovska}, {Nicastro}, {Noble},
  {Nowak}, {Primini}, {Siemiginowska}, {Smith}, \& {Wise}}]{fruscione2006}
{Fruscione}, A., {McDowell}, J.~C., {Allen}, G.~E., {Brickhouse}, N.~S.,
  {Burke}, D.~J., {Davis}, J.~E., {Durham}, N., {Elvis}, M., {Galle}, E.~C.,
  {Harris}, D.~E., {Huenemoerder}, D.~P., {Houck}, J.~C., {Ishibashi}, B.,
  {Karovska}, M., {Nicastro}, F., {Noble}, M.~S., {Nowak}, M.~A., {Primini},
  F.~A., {Siemiginowska}, A., {Smith}, R.~K., \& {Wise}, M. 2006, in Society of
  Photo-Optical Instrumentation Engineers (SPIE) Conference Series, Vol. 6270,
  Society of Photo-Optical Instrumentation Engineers (SPIE) Conference Series

\bibitem[{{Giacintucci} {et~al.}(2008){Giacintucci}, {Venturi}, {Macario},
  {Dallacasa}, {Brunetti}, {Markevitch}, {Cassano}, {Bardelli}, \&
  {Athreya}}]{giacintucci2008}
{Giacintucci}, S., {Venturi}, T., {Macario}, G., {Dallacasa}, D., {Brunetti},
  G., {Markevitch}, M., {Cassano}, R., {Bardelli}, S., \& {Athreya}, R. 2008,
  \aap, 486, 347

\bibitem[{{Giovannini} {et~al.}(2009){Giovannini}, {Bonafede}, {Feretti},
  {Govoni}, {Murgia}, {Ferrari}, \& {Monti}}]{giovannini2009}
{Giovannini}, G., {Bonafede}, A., {Feretti}, L., {Govoni}, F., {Murgia}, M.,
  {Ferrari}, F., \& {Monti}, G. 2009, \aap, 507, 1257

\bibitem[{{Hallman} \& {Markevitch}(2004)}]{hallman2004}
{Hallman}, E.~J., \& {Markevitch}, M. 2004, \apjl, 610, L81

\bibitem[{{Henning} {et~al.}(2009){Henning}, {Gantner}, {Burns}, \&
  {Hallman}}]{henning2009}
{Henning}, J.~W., {Gantner}, B., {Burns}, J.~O., \& {Hallman}, E.~J. 2009,
  \apj, 697, 1597

\bibitem[{{Henriksen} \& {Markevitch}(1996)}]{henriksen1996}
{Henriksen}, M.~J., \& {Markevitch}, M.~L. 1996, \apjl, 466, L79+

\bibitem[{{Hoeft} \& {Br{\"u}ggen}(2007)}]{hoeft2007}
{Hoeft}, M., \& {Br{\"u}ggen}, M. 2007, \mnras, 375, 77

\bibitem[Hudson et 
al.(2010)]{hudson2010} Hudson, D.~S., Mittal, R., Reiprich, T.~H., et al.\ 2010, \aap, 513, A37 

\bibitem[{{Kaastra}(1992)}]{Kaastra1992}
{Kaastra}, J.~S. 1992, An X-Ray Spectral Code for Optically Thin Plasmas
  (Internal SRON-Leiden Report, updated version 2.0)

\bibitem[{{Kang} {et~al.}(2012){Kang}, {Ryu}, \& {Jones}}]{kang2012}
{Kang}, H., {Ryu}, D., \& {Jones}, T.~W. 2012, \apj, 756, 97

\bibitem[{{Kempner} \& {Sarazin}(2001)}]{kempner2001}
{Kempner}, J.~C., \& {Sarazin}, C.~L. 2001, \apj, 548, 639

\bibitem[{{Kempner} {et~al.}(2003){Kempner}, {Sarazin}, \&
  {Markevitch}}]{kempner2003}
{Kempner}, J.~C., {Sarazin}, C.~L., \& {Markevitch}, M. 2003, \apj, 593, 291

\bibitem[{{Kurtz} {et~al.}(1992){Kurtz}, {Mink}, {Wyatt}, {Fabricant},
  {Torres}, {Kriss}, \& {Tonry}}]{Kurtz1992}
{Kurtz}, M.~J., {Mink}, D.~J., {Wyatt}, W.~F., {Fabricant}, D.~G., {Torres},
  G., {Kriss}, G.~A., \& {Tonry}, J.~L. 1992, in Astronomical Society of the
  Pacific Conference Series, Vol.~25, Astronomical Data Analysis Software and
  Systems I, ed. D.~M. {Worrall}, C.~{Biemesderfer}, \& J.~{Barnes}, 432--+

\bibitem[{{Liedahl} {et~al.}(1995){Liedahl}, {Osterheld}, \&
  {Goldstein}}]{liedahl1995}
{Liedahl}, D.~A., {Osterheld}, A.~L., \& {Goldstein}, W.~H. 1995, \apjl, 438,
  L115

\bibitem[{{Macario} {et~al.}(2011){Macario}, {Markevitch}, {Giacintucci},
  {Brunetti}, {Venturi}, \& {Murray}}]{marcario2011}
{Macario}, G., {Markevitch}, M., {Giacintucci}, S., {Brunetti}, G., {Venturi},
  T., \& {Murray}, S.~S. 2011, \apj, 728, 82

\bibitem[Markevitch(2010)]{markevitch2010} Markevitch, M.\ 2010, 
arXiv:1010.3660 

\bibitem[{{Markevitch} {et~al.}(2002){Markevitch}, {Gonzalez}, {David},
  {Vikhlinin}, {Murray}, {Forman}, {Jones}, \& {Tucker}}]{markevitch2002}
{Markevitch}, M., {Gonzalez}, A.~H., {David}, L., {Vikhlinin}, A., {Murray},
  S., {Forman}, W., {Jones}, C., \& {Tucker}, W. 2002, \apjl, 567, L27

\bibitem[{{Markevitch} {et~al.}(2005){Markevitch}, {Govoni}, {Brunetti}, \&
  {Jerius}}]{markevitch2005}
{Markevitch}, M., {Govoni}, F., {Brunetti}, G., \& {Jerius}, D. 2005, \apj,
  627, 733

\bibitem[{{Markevitch} {et~al.}(2000){Markevitch}, {Ponman}, {Nulsen}, {Bautz},
  {Burke}, {David}, {Davis}, {Donnelly}, {Forman}, {Jones}, {Kaastra},
  {Kellogg}, {Kim}, {Kolodziejczak}, {Mazzotta}, {Pagliaro}, {Patel}, {Van
  Speybroeck}, {Vikhlinin}, {Vrtilek}, {Wise}, \& {Zhao}}]{markevitch2000}
{Markevitch}, M., {Ponman}, T.~J., {Nulsen}, P.~E.~J., {Bautz}, M.~W., {Burke},
  D.~J., {David}, L.~P., {Davis}, D., {Donnelly}, R.~H., {Forman}, W.~R.,
  {Jones}, C., {Kaastra}, J., {Kellogg}, E., {Kim}, D.-W., {Kolodziejczak}, J.,
  {Mazzotta}, P., {Pagliaro}, A., {Patel}, S., {Van Speybroeck}, L.,
  {Vikhlinin}, A., {Vrtilek}, J., {Wise}, M., \& {Zhao}, P. 2000, \apj, 541,
  542

\bibitem[{{Markevitch} {et~al.}(1999){Markevitch}, {Sarazin}, \&
  {Vikhlinin}}]{markevitch1999}
{Markevitch}, M., {Sarazin}, C.~L., \& {Vikhlinin}, A. 1999, \apj, 521, 526

\bibitem[{{Markevitch} \& {Vikhlinin}(2007)}]{markevitch2007}
{Markevitch}, M., \& {Vikhlinin}, A. 2007, \physrep, 443, 1

\bibitem[{{Mastropietro} \& {Burkert}(2008)}]{mastropietro2008}
{Mastropietro}, C., \& {Burkert}, A. 2008, \mnras, 389, 967

\bibitem[Mazzotta et al.(2011)]{mazzotta2011} Mazzotta, P., Bourdin, 
H., Giacintucci, S., Markevitch, M., \& Venturi, T.\ 2011, \memsai, 82, 495 

\bibitem[{{Mewe} {et~al.}(1985){Mewe}, {Gronenschild}, \& {van den
  Oord}}]{mewe1985}
{Mewe}, R., {Gronenschild}, E.~H.~B.~M., \& {van den Oord}, G.~H.~J. 1985,
  \aaps, 62, 197

\bibitem[{{Mewe} {et~al.}(1986){Mewe}, {Lemen}, \& {van den Oord}}]{mewe1986}
{Mewe}, R., {Lemen}, J.~R., \& {van den Oord}, G.~H.~J. 1986, \aaps, 65, 511

\bibitem[{{Mink} {et~al.}(2007){Mink}, {Wyatt}, {Caldwell}, {Conroy}, {Furesz},
  \& {Tokarz}}]{mink2007}
{Mink}, D.~J., {Wyatt}, W.~F., {Caldwell}, N., {Conroy}, M.~A., {Furesz}, G.,
  \& {Tokarz}, S.~P. 2007, in Astronomical Society of the Pacific Conference
  Series, Vol. 376, Astronomical Data Analysis Software and Systems XVI, ed.
  {R.~A.~Shaw, F.~Hill, \& D.~J.~Bell}, 249--+

\bibitem[{{Okabe} \& {Umetsu}(2008)}]{okabe2008}
{Okabe}, N., \& {Umetsu}, K. 2008, \pasj, 60, 345

\bibitem[{{Owers} {et~al.}(2013){Owers}, {Baldry}, {Bauer}, {Bland-Hawthorn},
  {Brown}, {Cluver}, {Colless}, {Driver}, {Edge}, {Hopkins}, {van Kampen},
  {Lara-Lopez}, {Liske}, {Loveday}, {Pimbblet}, {Ponman}, \&
  {Robotham}}]{owers2013}
{Owers}, M.~S., {Baldry}, I.~K., {Bauer}, A.~E., {Bland-Hawthorn}, J., {Brown},
  M.~J.~I., {Cluver}, M.~E., {Colless}, M., {Driver}, S.~P., {Edge}, A.~C.,
  {Hopkins}, A.~M., {van Kampen}, E., {Lara-Lopez}, M.~A., {Liske}, J.,
  {Loveday}, J., {Pimbblet}, K.~A., {Ponman}, T., \& {Robotham}, A.~S.~G. 2013,
  \apj, 772, 104

\bibitem[{{Owers} {et~al.}(2009){Owers}, {Nulsen}, {Couch}, \&
  {Markevitch}}]{owers2009c}
{Owers}, M.~S., {Nulsen}, P.~E.~J., {Couch}, W.~J., \& {Markevitch}, M. 2009,
  \apj, 704, 1349

\bibitem[{{Owers} {et~al.}(2011){Owers}, {Randall}, {Nulsen}, {Couch}, {David},
  \& {Kempner}}]{owers2011a}
{Owers}, M.~S., {Randall}, S.~W., {Nulsen}, P.~E.~J., {Couch}, W.~J., {David},
  L.~P., \& {Kempner}, J.~C. 2011, \apj, 728, 27

\bibitem[{{Randall} {et~al.}(2008{\natexlab{a}}){Randall}, {Nulsen}, {Forman},
  {Jones}, {Machacek}, {Murray}, \& {Maughan}}]{randall2008}
{Randall}, S., {Nulsen}, P., {Forman}, W.~R., {Jones}, C., {Machacek}, M.,
  {Murray}, S.~S., \& {Maughan}, B. 2008{\natexlab{a}}, \apj, 688, 208

\bibitem[{{Randall} {et~al.}(2008{\natexlab{b}}){Randall}, {Markevitch},
  {Clowe}, {Gonzalez}, \& {Brada{\v c}}}]{randall2008b}
{Randall}, S.~W., {Markevitch}, M., {Clowe}, D., {Gonzalez}, A.~H., \&
  {Brada{\v c}}, M. 2008{\natexlab{b}}, \apj, 679, 1173

\bibitem[{{Rines} {et~al.}(2012){Rines}, {Geller}, {Diaferio}, \&
  {Kurtz}}]{rines2012}
{Rines}, K., {Geller}, M.~J., {Diaferio}, A., \& {Kurtz}, M.~J. 2012, ArXiv
  e-prints

\bibitem[{{Roediger} {et~al.}(2013){Roediger}, {Kraft}, {Forman}, {Nulsen}, \&
  {Churazov}}]{roediger2013}
{Roediger}, E., {Kraft}, R.~P., {Forman}, W.~R., {Nulsen}, P.~E.~J., \&
  {Churazov}, E. 2013, \apj, 764, 60

\bibitem[{{Rudnick} \& {Lemmerman}(2009)}]{rudnick2009}
{Rudnick}, L., \& {Lemmerman}, J.~A. 2009, \apj, 697, 1341

\bibitem[{{Russell} {et~al.}(2012){Russell}, {McNamara}, {Sanders}, {Fabian},
  {Nulsen}, {Canning}, {Baum}, {Donahue}, {Edge}, {King}, \&
  {O'Dea}}]{russell2012}
{Russell}, H.~R., {McNamara}, B.~R., {Sanders}, J.~S., {Fabian}, A.~C.,
  {Nulsen}, P.~E.~J., {Canning}, R.~E.~A., {Baum}, S.~A., {Donahue}, M.,
  {Edge}, A.~C., {King}, L.~J., \& {O'Dea}, C.~P. 2012, \mnras, 423, 236

\bibitem[{{Russell} {et~al.}(2010){Russell}, {Sanders}, {Fabian}, {Baum},
  {Donahue}, {Edge}, {McNamara}, \& {O'Dea}}]{russell2010}
{Russell}, H.~R., {Sanders}, J.~S., {Fabian}, A.~C., {Baum}, S.~A., {Donahue},
  M., {Edge}, A.~C., {McNamara}, B.~R., \& {O'Dea}, C.~P. 2010, \mnras, 406,
  1721

\bibitem[{{Sanders} \& {Fabian}(2001)}]{sanders2001}
{Sanders}, J.~S., \& {Fabian}, A.~C. 2001, \mnras, 325, 178

\bibitem[{{Sarazin}(2008)}]{sarazin2008}
{Sarazin}, C.~L. 2008, in Lecture Notes in Physics, Berlin Springer Verlag,
  Vol. 740, A Pan-Chromatic View of Clusters of Galaxies and the Large-Scale
  Structure, ed. M.~{Plionis}, O.~{L{\'o}pez-Cruz}, \& D.~{Hughes}, 1--4020

\bibitem[{{Serra} \& {Diaferio}(2013)}]{serra2013}
{Serra}, A.~L., \& {Diaferio}, A. 2013, \apj, 768, 116

\bibitem[{{Springel} \& {Farrar}(2007)}]{springel2007}
{Springel}, V., \& {Farrar}, G.~R. 2007, \mnras, 380, 911

\bibitem[{{van Weeren} {et~al.}(2011{\natexlab{a}}){van Weeren}, {Br{\"u}ggen},
  {R{\"o}ttgering}, {Hoeft}, {Nuza}, \& {Intema}}]{vanweeren2011}
{van Weeren}, R.~J., {Br{\"u}ggen}, M., {R{\"o}ttgering}, H.~J.~A., {Hoeft},
  M., {Nuza}, S.~E., \& {Intema}, H.~T. 2011{\natexlab{a}}, \aap, 533, A35

\bibitem[{{van Weeren} {et~al.}(2011{\natexlab{b}}){van Weeren}, {Hoeft},
  {R{\"o}ttgering}, {Br{\"u}ggen}, {Intema}, \& {van Velzen}}]{vanweeren2011a}
{van Weeren}, R.~J., {Hoeft}, M., {R{\"o}ttgering}, H.~J.~A., {Br{\"u}ggen},
  M., {Intema}, H.~T., \& {van Velzen}, S. 2011{\natexlab{b}}, \aap, 528, A38

\bibitem[{{van Weeren} {et~al.}(2010){van Weeren}, {R{\"o}ttgering},
  {Br{\"u}ggen}, \& {Hoeft}}]{vanweeren2010}
{van Weeren}, R.~J., {R{\"o}ttgering}, H.~J.~A., {Br{\"u}ggen}, M., \& {Hoeft},
  M. 2010, Science, 330, 347

\bibitem[{{Vikhlinin} {et~al.}(2001{\natexlab{a}}){Vikhlinin}, {Markevitch}, \&
  {Murray}}]{vikhlinin2001b}
{Vikhlinin}, A., {Markevitch}, M., \& {Murray}, S.~S. 2001{\natexlab{a}}, \apj,
  551, 160

\bibitem[{{Vikhlinin} {et~al.}(2001{\natexlab{b}}){Vikhlinin}, {Markevitch}, \&
  {Murray}}]{vikhlinin2001a}
---. 2001{\natexlab{b}}, \apjl, 549, L47

\bibitem[{{Vikhlinin} \& {Markevitch}(2002)}]{vikhlinin2002}
{Vikhlinin}, A.~A., \& {Markevitch}, M.~L. 2002, Astronomy Letters, 28, 495

\end{thebibliography}
\end{document}